\documentclass{svmult}
\usepackage[]{natbib}
\newcommand{\posscitet}[1]{\citeauthor{#1}'s (\citeyear{#1})}

\usepackage{xspace}
\usepackage{filecontents}
\usepackage{graphicx}
\usepackage{subfigure}
\usepackage{amsmath}

\usepackage{url}
\urldef\AschwandenURL\url{http://download.springer.com/static/pdf/113/art%253A10.1007%252Fs11214-014-0054-6.pdf?auth66=1412352067_5e2b75759c41abad274561cc8078b30f&ext=.pdf}

\usepackage{mathptmx}       
\usepackage{helvet}         
\usepackage{courier}        
\usepackage{type1cm}        
\usepackage{makeidx}         
\usepackage{graphicx}        
\usepackage{multicol}        
\usepackage[bottom]{footmisc}

\makeindex             

\newcommand{\latin}[1]{{\it #1}}
\newcommand{\ie}{\latin{i.e.}\@\xspace}
\newcommand{\eg}{\latin{e.g.}\@\xspace}

\newcommand{\etc}{\latin{etc.}\@\xspace}

\newcommand{\quotebracketed}[1]{[#1]}

\def\signed #1{{\leavevmode\unskip\nobreak\hfil\penalty50\hskip2em%
\hbox{}\nobreak\hfil{#1}%
    \parfillskip=0pt \finalhyphendemerits=0 \endgraf}}

\newcommand{\citedquotesrc}{}
\newenvironment{citedquote}[2]{%
\renewcommand{\citedquotesrc}{\citep[][#2]{#1}}%
\begin{quote}}{%
\signed{\citedquotesrc}\end{quote}%
}

\newcommand{\slabel}[1]{\label{sec:#1}}

\newcommand{\Sref}[1]{Section~\ref{sec:#1}}

\newcommand{\flabel}[1]{\label{fig:#1}}

\newcommand{\Fref}[1]{Figure~\ref{fig:#1}}

\begin{document}
\title*{25 Years of Self-Organized Criticality: Concepts and Controversies}
\author{Nicholas W. Watkins\inst{1,2,3,4}\and
Gunnar Pruessner\inst{5}\and
Sandra C. Chapman\inst{6,7}\and
Norma B. Crosby\inst{8}\and
Henrik J. Jensen\inst{9}}

\institute{%
\inst{1}Max Planck Institute for the Physics of Complex Systems,
Dresden, Germany,
\email{NickWatkins@mykolab.com}
\and
\inst{2}London School of Economics and Political Science,
London, United Kingdom
\and
\inst{3} Open University, Milton Keynes, United Kingdom
\and
\inst{4}University of Warwick,
Coventry,
United Kingdom
\and
\inst{5}Imperial College London,
London,
United Kingdom,
\email{g.pruessner@imperial.ac.uk}
\and
\inst{6}University of Warwick
Coventry,
United Kingdom,
\email{s.c.chapman@warwick.ac.uk}
\and 
\inst{7} UiT The Arctic University of Norway, Tromso, Norway
\and 
\inst{8}Belgian Institute for Space Aeronomy,
Brussels,
Belgium,
\email{Norma.Crosby@aeronomie.be}
\and
\inst{9}Imperial College London,
London,
United Kingdom,
\email{h.jensen@imperial.ac.uk}
}
\date{\date}
\authorrunning{Watkins, Pruessner, Chapman, Crosby, Jensen}

\maketitle

\abstract{Introduced by the late Per Bak and his colleagues, self-organized criticality (SOC) has been one of
the most stimulating concepts to come out of statistical mechanics and
condensed matter theory in the last few decades, and has played a significant role in the development of complexity science. SOC, and more generally fractals and power laws, have attracted much comment, ranging from the very positive to the  polemical.  The other papers \citep{AschwandenETAL:2014,McAteerETAL:2014,SharmaETAL:2015}  in this special issue 
showcase the considerable body of observations in solar, magnetospheric and
fusion plasma inspired by the SOC idea, and expose the fertile role
the new paradigm has played in approaches to modeling and understanding multiscale plasma
instabilities. This very broad impact, and the
necessary process of adapting a scientific hypothesis to the conditions
of a given physical system, has meant that SOC as studied in these
fields has sometimes differed significantly from the definition originally
given by its creators. In Bak's own field of theoretical physics there
are significant observational and theoretical open questions, even 25
years on \citep{PruessnerBook}. One aim of the present review is to address the dichotomy between the great reception SOC has received in some areas, and its shortcomings, as they became manifest
 in the controversies it triggered. Our article tries to clear up what we think are  misunderstandings of SOC in fields more remote from its origins in statistical mechanics,
 condensed matter and dynamical systems by revisiting Bak, Tang and
 Wiesenfeld's  original papers.}

 \pagebreak

\begin{citedquote}{Anderson:2011}{p.~112}
The idea of self-organized criticality seems to me to be, not the right
and unique solution to these and other  similar problems, but to have
paradigmatic value, as the kind of generalization which will
characterize the next stage of physics. \quotebracketed{\dots} In the 21st century one revolution which can take place is the construction of generalizations which jump and jumble the hierarchies, or generalizations which allow scale-free or scale-transcending phenomena. The paradigm for the first is broken symmetry, for the second self-organized criticality.
\end{citedquote}

\begin{citedquote}{CoenCoen:1998}{}
This is a very complicated case, Maude. You know, a lotta ins, lotta outs, lotta what-have-you's. And, uh, lotta strands to keep in my head, man. Lotta strands in old Duder's head.
\end{citedquote}

\section{Introduction and synopsis}\slabel{introduction}

The late Per Bak's concept  of Self-Organized
Criticality (SOC), first stated in his seminal papers with Chao Tang and
Kurt Wiesenfeld \citep*[][in the following abbreviated to BTW]{BTW87,BTW88}
has been extremely stimulating, with over 6600 citations since
\citeyear{BTW87}.  SOC continues to live a
number of parallel lives in various fields, such as statistical
mechanics, seismology \citep[\eg][]{BakETAL:2002a}, materials science
\citep[\eg][]{AltshulerETAL:2001}, condensed matter theory
\citep[\eg][]{WijngaardenETAL:2006}, ecology
\citep[\eg][]{MalamudMoreinTurcotte:1998}, evolution
\citep[\eg][]{SneppenETAL:1995}, high energy astrophysics
\citep[\eg][]{NegoroETAL:1995,DendyETAL:1998,MineshigeNegoro:1999,AudardETAL:2000}, neuroscience
\citep[\eg][]{BedardKroegerDestexhe:2006,Chialvo:2010} and sociology
\citep[\eg][]{Hoffmann:2005}. 
 
In particular, SOC has become a research field in laboratory fusion
plasmas, solar physics and magnetospheric physics, reviewed in the
complementary papers \citep{AschwandenETAL:2014,McAteerETAL:2014,SharmaETAL:2015} in
this volume.  Like them, our own paper results from two workshops at the
International Space Science Institute in 2012 and 2013 (\url{http://www.issibern.ch/teams/s-o-turbulence/}).

Despite its success, however, SOC has often divided opinions, even among
experts. It has attracted significant criticism
\citep[\eg][]{PerkovicDahmenSethna:1995,Krommes:2000,Frigg:2003,StumpfPorter:2012},
some of it deserved, some of it polemical,  to such a degree that some
authors in condensed matter physics will avoid mentioning it altogether
\citep[\eg][]{AlvaradoETAL:2013}. Although numerous reviews
\citep{Turcotte:1999,AlstromETAL:2004,Sornette:2006,Dhar:2006}
and several book-length surveys of theory 
\citep{JensenBook,PruessnerBook,ChristensenMoloney:2005} and applications \citep{Hergarten:2002,Aschwanden:2011,Rodriguez-Iturbe:2001} exist,  
the enduring ``hectic air of controversy''  \citep[][p.~125]{JensenBook}
  has ensured
that many people remain uncertain both of SOC's long term status, and of
its net contribution to science. This will undoubtedly also be true for
many readers of Space Science Reviews, browsing the group of papers
centered on SOC in space and plasma physics that are collected in this
volume.

Our contribution to the present volume aims to both complement the surveys of SOC
in space and lab plasmas in the accompanying papers and to address this
controversy. We first, in  \Sref{Paradigm}, discuss why   the
multiscale avalanching paradigm, of which SOC is the best known example,
is relevant to both space and laboratory plasmas. We then, in
\Sref{WhichSoc},  clarify just what kind of ``SOC'' we are talking about
in this paper, by distinguishing as briefly as we can between the main
different perceptions  of SOC (see \Fref{soc_diagram}) that one can find in various
research disciplines, including space  plasma physics.  The first of
these pictures is BTW SOC, the SOC that was introduced in BTW's original
papers. It  has a theoretical core  underpinning it which not only
remains essentially intact but also has been substantially clarified
over 25 years, and so it is the SOC which we discuss in the rest of the
paper.  

We thus continue by re-examining BTW SOC's
foundations. We do so by reference to, and quotes from, the key original
papers, in search of BTW's original claims, in order to understand the
interpretations (and misinterpretations) which have been made of them.
We first revisit BTW's motivation for introducing  the SOC idea, which
they stated most clearly in \citep*{BakChen:1989}. In   \Sref{BTWmotivation},
\Sref{BTWpostulate}, and \Sref{BTWcritical}, closely following this key\footnote{and relatively little-known compared to the BTW papers of 1987-88} paper,  we will recap the reasoning which led BTW to their postulate, showing how it does in fact give a relatively precise definition of SOC.

\Sref{phenotypegenotype} then brings us up to date by linking the
preceding discussion to the contemporary literature on SOC. In so doing
we identify the necessary and sufficient conditions for SOC. Not
differentiating necessary and sufficient conditions is, we believe, one
source of the erroneous beliefs (sometimes found in the literature)
that everything that is avalanching must be critical and
self-organized, or, conversely, that everything that displays
long-ranged correlations or a power law must be an instance of
(self-organized) criticality.
 
These errors in logic, as well as a loose interpretation of BTW's core
idea have helped to create the divergence of versions of SOC noted above
and some of the controversy that has tended to surround SOC. In
\Sref{controversy} we address some of the most prominent of these,
giving our own views  about some misconceptions and subsequent
conflicts that have arisen. This section is an opinion piece insofar as
we will try to point out and clarify certain issues, which  we
think  have caused unnecessary problems in the past. 

We then balance our discussion of controversy by noting that in several areas of science SOC has been a success story.   \Sref{SOCwild} thus discusses how  this has happened, in a  context most relevant to our paper. We briefly discuss how SOC has provided a paradigm for, and thereby consolidation of, existing observations which lacked context, in solar, magnetospheric and fusion plasma physics.  Finally, we offer some concluding remarks and perspectives on future research in \Sref{conclusion}. 

Appendix A discusses, in a fairly self-contained manner, the more
technical topic of scaling, intended to
complement our discussion of SOC by setting the BTW worldview, SOC and
its foundation in some of
its broader theoretical context. It may be omitted in a first reading. 
In it  we discuss the general ideas of scaling, which
underpin a whole range of disciplines  
 such as critical phenomena and dynamical systems, as well as methods, like dimensional analysis  \citep[\eg][]{Buckingham:1914}. 

Our team of authors has a somewhat unique perspective in that our
previous involvement with SOC and complexity ranges from  condensed
matter theory \citep{JensenBook,PruessnerBook}, where the SOC concept originated, to
solar astrophysics \citep{CrosbyAschwandenDennis:1993,WatkinsChapmanRosenberg:2009}, complex  plasma physics both in the   coupled, magnetosphere and turbulent solar wind \citep{Chapman:1998,Watkins:1999,LuiETAL:2000,Chapman:2001,Watkins:2001,Freeman:2002,Watkins:2002} and in fusion reactors \citep{ChapmanETAL:2001,DendyChapmanPaczuski:2007},
and SOC-inspired (or informed) cross-disciplinary complexity research in the environmental sciences \citep{Watkins:2008,Watkins:2013,GravesETAL:2014}. We have thus tried to cater both
for readers interested  in the theoretical foundations of SOC and those
concerned with its applications to nature. We hope we have clarified
that SOC remains a strong, relevant, scientific theory, even if it is
not always ``how Nature works''.
 
Some readers will, rightly, read parts of this contribution as an opinion piece;
however, we have tried to support our views by quotes and references
wherever possible. Readers with a background in statistical mechanics
may be interested in the historical context that led to the development
of SOC, in particular
Secs.~\ref{sec:BTWmotivation}--\ref{sec:phenotypegenotype}, but also
\Sref{controversy} for the controversies surrounding SOC. Those from the
plasma physics community will probably be interested in
Secs.~\ref{sec:Paradigm}, \ref{sec:BTWpostulate}, 
\ref{sec:phenotypegenotype}, \ref{sec:controversy} and \ref{sec:SOCwild} with a focus on
applications in plasma physics. 
Those who are mostly interested in the controversies surrounding SOC
will benefit particularly from reading
Secs.~\ref{sec:WhichSoc}--\ref{sec:BTWcritical} and
\ref{sec:controversy}. We obviously believe that the other sections, in
particular the conclusions in \Sref{conclusion}, are of broad interest and
certainly worth reading, but most of the sections are fairly
self-contained, inviting the reader to make their own selection. We have tried to facilitate this by prefacing the most technical sections by bullet point summaries of what they contain.
 
\section{Why is multiscale avalanching a relevant paradigm for plasmas?}
\slabel{Paradigm}
Any new paradigm prompts experimental testing and examination. In
the Popperian account of science  \citep{Popper:1963}, this process is idealized as one of
theoretical conjecture and experimental refutation. The living process
of science often needs to be more pragmatic \citep{Ziman:2002}. New theoretical conjectures can prompt
ordering of observations in new ways which in turn can lead to new
insights. In this more pragmatic view, new observational insights can
stand in their own right, independent of the original theoretical
conjecture, indeed, the original conjecture can be heavily modified, or
ultimately abandoned. The pragmatic approach is particularly prevalent
in the sciences of observed natural systems typified by geology, seismology and astronomy, where there are limitations on the possibility of controlled experiments. Experimental refutation, or
indeed, confirmation, is often, to some extent, supplemented by a search for order and pattern in a
broad, inhomogeneous collection of observed phenomena. Theoretical
insight then plays a key role in informing the patterns for which we
search in the data. Arguably, SOC has played such a role in the observed
phenomenology of plasmas both in astrophysics and in the laboratory.

Over the last two decades (see e.g. \cite{DendyHelander:1997,PerroneETAL:2013} and references therein)  it has been recognized that key emergent phenomenology in confined tokamak plasma experiments for fusion cannot
be explained from the physics of plasmas that are quasilinear, or that
are close to equilibrium. These confined plasmas show transitions
between multiple, metastable states, and reconfiguration and transport
and release of energy in a bursty, intermittent manner. These plasmas
support multiple possible instabilities which can lead to energy
release, while the energy release itself occurs on multiple scales. The
underlying physics of individual instabilities, and of collective
nonlinear phenomena such as that of structure formation, turbulence and
reconnection, is common across astrophysical plasmas. Furthermore,
astrophysics offers natural confined plasmas: planetary magnetospheres
with either intrinsic or induced confining magnetic fields, and stars,
where the confinement is gravitational. These confined plasmas are
particularly well diagnosed in our own solar system, and these much larger,
astronomical systems support a sufficient range of
spatio-temporal scales that the statistics of burst sizes can be
resolved. The availability of the paradigm of SOC has led to
observational testing \citep{LuiETAL:2000,Uritsky:2002} of the conjecture
 that these burst
sizes are  distributed as  power laws both in solar flares \citep{CrosbyAschwandenDennis:1993,Aschwanden:2011}, and in the
detailed morphology of earth's aurora \citep{Chapman:1998}.

All of these systems, share common features, they are driven,
dissipating and far from equilibrium, releasing energy in a bursty
intermittent manner on multiple scales, and there are many detailed
routes to instability that can lead to this energy release and
reconfiguration. In SOC parlance, one might describe this as
``multiscale avalanching''. Furthermore, detailed event analysis of the
onset of energy release, for example in solar flares, or substorms in
the earth's magnetosphere, reveals that the threshold to instability is
slowly approached under driving, and the subsequent reconfiguration is
fast. Implicitly there is a separation of timescales --- slow driving
and fast relaxation, which we will see below is typical of SOC systems. 

Practically, one can consider two approaches to the physics of such a
system. One is to address the detailed plasma physics of an energy
release event in isolation. The relevant questions are then, what is the
cause of the onset of the instability? What is the specific instability
mechanism? The approach is by modeling and numerical simulations, \ie
solving the fullest tractable set of plasma physics equations, and
comparison to (suitably selected) observed events against which to test
the theory. This approach is best suited to physics which occurs on {\em
one} specific spatio-temporal scale. 

Alternatively, one can consider the situation where the physics on {\em
all} scales is equally important, and is, furthermore, strongly coupled.
Homogeneous turbulence is a classic example of such a process, though distinct from SOC \citep{ChapmanWatkins:2009,ChapmanEA:2009}. In this
case the ``bottom up'' approach of solving for the dynamics of
individual events is intractable. Instead, one can look to the success
of the renormalization group approach \citep{Wilson:1971a,Wilson:1979}
in critical phenomena. Here, one needs to characterize the fundamental
local interaction, and how it coarse-grains as more and more elements in
the system are aggregated. Central to the structure of such a model is
self-similar scaling (the system looks the same on all scales subject to
a rescaling), leading to power law distributions of (event) sizes and
power law (long-range) correlations as the key observable. Importantly,
a broad range of different detailed, microscopic interactions, on
coarse-graining, lead to the same collective behavior, thus one expects
the same essential phenomenology to be ubiquitous.

The classic triumphs of the renormalization group (RG) in critical phenomena
\citep{Wilson:1971a,Wilson:1979} have been for systems in equilibrium.
Extensions of the RG approach to non-equilibrium, either relaxing to equilibrium or staying
far from it, have been developed successfully over the last 40 years or
so (e.g \cite{ChangETAL:1992}), but a successful application to plasma physics remains elusive (though see for example \cite{Balescu:1997}).
It is for these diverse plasma systems that SOC offers
considerable attractions as a paradigm.
As we will see in \Sref{BTWcritical}, SOC introduces dynamics by
enforcing a separation of time scales, \ie the build-up to instability is
slow, while relaxation is fast. This fast relaxation leads to
avalanche-like, bursty energy release on a broad range of scales. The
dynamics of an avalanche is fundamentally multiscale, it occurs by
coupling across many spatial scales in the system. Importantly, the
statistics of energy release events, indeed, the dynamics, are not
sensitive to the details of the instability, thus in a plasma where many
instabilities and routes to instability are possible, one expects to see
the same, robust emergent behavior. Indeed, one could identify a
paradigm for SOC in plasmas, or perhaps more accurately, ``multiscale
avalanching'', based on these properties alone, which are sufficient to
provide a new, insightful framework for ordering the observations. 

\section{Perceptions and receptions of SOC}
\slabel{WhichSoc}

The interaction of BTW's papers and their many readers has led to 
nested\footnote{Although nested, we accept that at least the
outermost region, clearly the boldest claim, is not necessarily a
superset of the claims inside.}  perceptions of SOC, as
illustrated in \Fref{soc_diagram}. We can summarize these as essentially
four, in order of increasing ubiquitousness: 
\begin{itemize}
  \item {\em Self-tuned phase transitions can (and do) exist in nature}
  --- The core idea of SOC, clearly enunciated by  \cite{BakChen:1989},
  which was presented as a dynamical origin of   spatio-temporal fractals
  in nature\footnote{Interestingly \citet[][see also \citealp{Milovanov:2013}]{MilovanovIomin:2014}
have asserted that in some contexts the core idea of SOC as a self-tuned phase transition can be further demonstrated using their topological approach. They used a backbone map onto a Cayley tree, and the formalism of the discrete Anderson nonlinear Schroedinger equation (DANSE). The DANSE has been used to describe physical systems including Bose-Einstein condensates and arrays of nonlinear waveguides, but its theoretical interest is more far-reaching, because it ``also serves as a
paradigmatic model for a wide class of physical problems
where interplay of nonlinearity and disorder is important" \citep{PikovskyShepelyansky:2008}.}.
  \item {\em All fractals in nature are caused by SOC} --- A much more
  sweeping claim, but one which a reader could have been forgiven for
  inferring from the abstract of the same paper \citep{BakChen:1989}.
  \item {\em All power laws are caused by SOC} --- An even more sweeping
  claim, never to our knowledge made by Bak, but which many readers
  might easily have inferred from reading  the discussion of  Zipf's
  law in the first chapter of his book  \citep{Bak:1996}.
  \item {\em The contingency of nature is caused by SOC} --- See for
  example the abstract of  \citep{BakPaczuski:1995}, and
  \posscitet{Bak:2000} review of \cite{Buchanan:2000}.
\end{itemize}

\begin{figure}[t] 
\begin{center}
\includegraphics{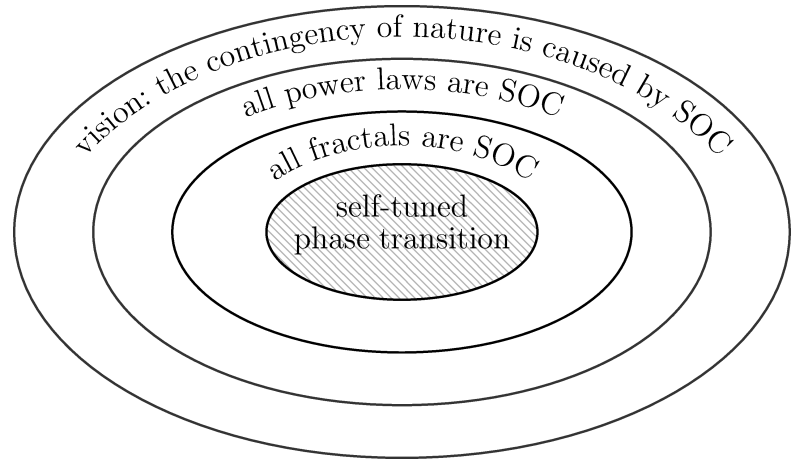}
\end{center}
\caption{\flabel{soc_diagram}
A schematic representation of the range of perceptions of SOC in the
literature,  from the most minimal at the center to the most visionary.
As explained in the text, the hatched core is the proposal that a 
mechanism exists in nature whereby some systems tune \emph{themselves} to a phase
transition. This mechanism has sometimes been promoted as the primary (or even single)
cause of fractals in nature (second circle).  Some authors have regarded
fractals and power laws as synonymous (third circle), and proposed that SOC was needed as the underlying mechanism for the latter as well, despite the many alternative explanations in many cases. The outer-most region
considers contingency in nature as the signature of SOC.}
\end{figure} 

The clear divergence between these pictures of SOC, and the fact that
all of them have had at least some adherents, has had some important
consequences. An ever increasing diversity and confusion about claims,
proofs and evidence resulted in a muddled perception of the status 
of SOC; of how it explains natural phenomena; and which ones. To this day,
except for computational confirmation of the core claim, there is no
unambiguous, unquestioned evidence for any of the claims above, even
though they inspired much research. However, it is also fair to say that
experimental, observational, numerical and analytical work is homing in
to corroborate at least the core claim.

Whether correct or not in its generality, the first picture, the core
SOC idea, was from the outset relatively tightly defined,  being
formulated in the language of mathematical physics and condensed matter
theory. It was actively pursued and debated by these communities from
the outset, and has been the subject of several book length treatments
including those of two of the present authors
\citep{JensenBook,PruessnerBook}.

The second and third pictures have long been known to be
wrong\footnote{See e.g. the extensive discussions in
\citep{JensenBook,Sornette:2006}. Chapter 14 of the latter gives a
particularly useful comprehensive overview of power laws observed in
nature which are caused by processes that are fundamentally different
from SOC, and contrasts these with various mechanisms for SOC in the
accompanying Chapter 15.}, and yet have at the same time been widely
influential. Being at the same time a target for criticism and polemic,
and a source of creative misunderstanding, they provide an interesting
present-day example for historians and philosophers of science of how
error and miscommunication can sometimes have positive as well as
negative side effects.  

The fourth, and most visionary, picture essentially  expressed a new
paradigm for complexity science, which we may call ``complexity and
contingency from criticality'' as opposed to contingency from low dimensional chaos. Bak
put this vision clearly, late in his life, in  his review of
\cite{Buchanan:2000} \begin{citedquote}{Bak:2000}{56--57} The tool of
history, certainly, is story-telling after the fact. Why is this? Why
does it make no useful theoretical predictions? Why is it, in other
words, that ``Life is understood backwards, but must be lived
forwards'', as philosopher S{\o}ren Kierkegaard put it.  [\dots]

Buchanan wants us to know that we live in a special time in which new
ideas are beginning to make it possible to see why history is the way it
is. Surprisingly, perhaps, the ideas it uses find their origin not in
history, but in theoretical physics. He proposes to explain why history
is and even must be punctuated by dramatic, unpredictable upheavals. He
promotes a theory declaring that all past efforts to perceive cycles,
progressions and understandable patterns of change in history have
necessarily been doomed to failure. 

[\dots]

``Contingency is the affirmation of control by immediate events over
destiny, the kingdom lost for want of a horseshoe nail,'' as biologist
Stephen J. Gould has observed. And contingency is the hallmark of the
critical state.
\end{citedquote}

This fourth interpretation of SOC is arguably at least as interesting as
the first, though much more speculative. It is, however, of much less
importance to the astrophysical plasma context of the present paper. We will
thus concentrate entirely on the first, and will only discuss the second
and third erroneous perceptions when we need to explain how they have
confused the issue.

\section{BTW's Stated Motivation: ``a dynamical theory of the physics of  fractals''}
\slabel{BTWmotivation}
In this section we 
describe BTW's aims in proposing SOC, and how Bak and Chen in their 1989 Mandelbrot Festschrift paper \citep{BakChen:1989}:
\begin{itemize}
  \item Quoted the widespread evidence for spatial fractals, and power law spatial correlation functions, and echoed Kadanoff's call for a physical explanation,
  \item Remarked on the parallel unsolved time domain problem of ``1/f" noise
  \item Proposed the bold idea that natural complex systems could self-organize to a particular kind of state that produced these effects, analogous with those seen in laboratory ``critical" systems near a phase transition-hence Self Organized Criticality.
\end{itemize}

{\bf Spatial Fractals:} In their 1989 paper, Bak and Chen gave what we believe to be their clearest statement of the original SOC idea. They first noted the evidence for the widespread existence of spatial fractals:
\begin{citedquote}{BakChen:1989}{p.~5}
 The importance of Mandelbrot's discovery that fractals occur widespread
 in nature can hardly be exaggerated. Many things which we used to think
 of as messy and structureless are in fact characterized by well-defined
 power law spatial correlation functions. By now, we are so used to
 seeing fractals that we are tempted to feel that we understand them.
 But do we simply have to accept their existence as ``God-given''
 without further explanation or is it possible to construct a dynamical
 theory of the physics of fractals?
\end{citedquote}
It is important to note that the power laws of concern to Bak and Chen
were in the {\em correlations} between fluctuations in space, rather
than the general question of power law \emph{size} distributions
 in nature, a point we
will return to (\eg \Sref{phenotypegenotype}). BTW used power law
distribution functions as \emph{proxy} for power law correlations,
making that link explicit at an early stage \citep[][p.~369]{BTW88}. In
general, it is by no means  clear that size distributions with no
clear connection to spatial correlation (or avalanches), such as those
of fractured frozen potatoes  \citep{Oddershede:1993}, the distribution
of lunar crater sizes \citep{Head:2010}, or the length of queues in
Britain's National Health Service (and her pubs)
\citep{SmethurstWilliams:2001,FreckletonSutherland:2001}  would (or should) ever
have been seriously intended to be in the remit of BTW's SOC. We would
argue that it remains unhelpful to try to define a notion of ``SOC''
that is sufficiently elastic to encompass them.

{\bf Fractals in time and $1/f$ noise:} Bak and Chen went on to highlight the ubiquity of fractals in time: 
\begin{citedquote}{BakChen:1989}{p.~5}
There is another ubiquitous phenomenon which has defied explanation for
decades. The signal (water, electrical current, light, prices, \dots)
from a variety of sources has a power spectrum decaying with an exponent
near unity at low frequencies \dots This type of behavior is known as
``1/ f'' noise, or flicker noise.
\end{citedquote}
The ``$1/f$'' noise which BTW referred to was discovered by Schottky,
early in the 20th century. 
A $1/f$ power spectrum is generally regarded as the fingerprint of
(temporal) correlations so strong that any future state must be
considered a function of the system's entire history
\citep[][p.~9]{JensenBook}. This remains true for generalized $1/f$
spectra, \ie across a range of power law dependences of the power
spectrum on $f$.

It is important to realize that rather than
``defying'' explanation, it had in fact been the subject of many
explanations
\citep[\eg][]{vanderZiel:1950,SchickVerveen:1974,Weissman:1988}, but
that BTW found these unsatisfying and lacking in generality. In
magnetospheric physics the presence of ``$1/f$'' spectra in geomagnetic
indices and other ground-based magnetic measurements
\citep{Tsurutani:1990,Weatherwax:2000} was, early on, one of several key supporting pieces of evidence of SOC.

{\bf The SOC postulate:} 
A perceived need to unify the above two aspects of fractality, and, importantly, a
claimed absence of existing ways to do so, led BTW to postulate the idea
of SOC. Apparently, they were guided by the observation of scaling in
space and time (fractals and generalized $1/f$ noise) in equilibrium and
non-equilibrium critical phenomena, such as the Ising Model
\citep{Stanley:1971,HohenbergHalperin:1977}. Bak and Chen put it this
way:  
\begin{citedquote}{BakChen:1989}{p.~5}
Strangely enough, just as those working on fractal phenomena in nature
never seem to be interested in the temporal aspects of the phenomenon,
\quotebracketed{\dots} those working on ``1/ f'' noise never bother with
the spatial structure of the source of the signal. We believe that those
two phenomena are often two sides of the same coin: they are the spatial
and temporal manifestations of a self-organized critical state.
\end{citedquote}
Bak and Chen prefaced this (already very bold) claim  in the paper with
one of the most memorably terse abstracts in the history of science,
wich may be called the ``SOC postulate'':
{\em ``Fractals in nature originate from self-organized critical
dynamical processes''}, expanded on  by a comment on the first page where
they said:
\begin{citedquote}{BakChen:1989}{p.~5}
We see fractals as snapshots of systems operating at the self-organized
critical state.
\end{citedquote}

The  gap between the   relatively specific idea of explaining space-time
fractal avalanching phenomena and therefore spatio-temporal
correlations,\footnote{See, however, the discussion at the end of
\Sref{confusion}.} and the  aspiration that many perceived
to explain {\em any} fractal in space or time, or even any power law
distribution, has been a perennial problem, and a key source of the  controversy  and
misunderstandings that still surround SOC. 
 
\section{Self-organization, the SOC postulate and BTW's definition of SOC}
\slabel{BTWpostulate}
Having seen in the previous section what the problem (spacetime fractals) that BTW sought to solve was, we now will show in more detail how the solution (the SOC hypothesis) was meant to do so.
In the following section, we will summarize four key points about the scientific
program  followed by BTW when they
formulated SOC:
\begin{itemize}
  \item They argued that spatial and temporal scaling must usually be unavoidably connected
  \item They posited that in contrast to phase transitions (or chaos) seen at a fixed point in control parameter space there must be a more robust (and thus widespread)  new kind of spatiotemporal critical behaviour which resulted from self-organization and for which their sandpile was the exemplar (i.e. SOC)
  \item They identified conditions for SOC behavior to be seen in a system; later recast by Jensen as  `slowly driven interaction dominated and thresholded", and also highlighted the role of dissipation in maintaining such a state
  \item And they asserted that spacetime fractals were snapshots of the SOC state.
\end{itemize}

Rather than the impossibly broad, and with hindsight unnecessary, goal of explaining all power laws in
nature with one mechanism, a close rereading of \citep[][as well as
\citealp{BTW87,BTW88}, in which the SOC concept was
launched]{BakChen:1989} shows quite clearly that the aim of SOC was to
unify dynamically evolving spatial and temporal fractals. BTW were taking as a cue
\posscitet{Kadanoff:1986} famous question  ``Fractals: Where's the
Physics?'', which itself had been aimed at a fractal ``industry'' which was experiencing its first wave
of enthusiasm at that point. In  a volume of papers dedicated to Mandelbrot, \citet{BakChen:1989} responded equally
boldly and provocatively to Kadanoff that: ``Fractals in nature originate from
self-organized critical dynamical processes''.  Beyond the immediate
goal lay the even more ambitious one of accepting the challenges posed
by two Nobel Laureates: Phil \posscitet{Anderson:1972} celebrated essay ``More is different'' on
complexity science, and Ken \posscitet{Wilson:1979} invitation to a wider
adoption of what is arguably the most powerful tool in statistical
mechanics, the renormalization group. 

Firstly, they argued that spatial and temporal scaling were
intrinsically linked,
\ie that the scaling historically observed in time series as $1/f$ noise
\citep{vanderZiel:1950} is related to the spatial scaling that became
prominent with the advent of \posscitet{Mandelbrot:1983} fractals
\citep[\eg][p.~7]{Feder:1988}:
\begin{citedquote}{BakChen:1989}{p.~5}
Actually, for those (like us) who are brought up as condensed matter
physicists it is hard to believe that long-range spatial and temporal
correlations can exist independently. A local signal cannot be
``robust'' and remain coherent over long times in the presence of any
amount of noise, unless stabilized by the interactions with its
environment. And a large, coherent spatial structure cannot disappear (or
be created) instantly. For an illustration, think of the temporal
distribution of sunshine, which must be correlated with the spatial
distribution of clouds, through the dynamics of meteorology.
\end{citedquote}
It has been argued subsequently, however, that scaling in time is rather common
\citep{GrinsteinLeeSachdev:1990} in non-equilibrium, and even 
in equilibrium dynamics, which is otherwise ``a rotten place to hunt for
generic scale invariance'' \citep[][p.~262]{Grinstein:1995}. In other words, space
and time fractality need not in fact be related.  Prior knowledge that one is dealing with a spatially extended, and importantly, connected, system may make such a connection more likely.  The precise way to check, at least in principle, is to measure a spatiotemporal correlation function 
\begin{equation}
C(r,t) = <f(r_0,t_0)f(r_0+r,t_0+t)> - <f>^2 
\end{equation}
and check if one has scaling (i.e. algebraic rather than exponential dependence) in {\em both} $r$ and $t$.

It is interesting with hindsight that as early as 1967 Mandelbrot had realised that scaling in time, and thus ``1/f" noise, need not always be attributed to the kind of stationary long range dependence seen in his own fractional Gaussian noise models. An alternative model, which was  only (in his words) ``conditionally" stationary, was the fractional renewal process he discussed in \citep{Mandelbrot:1967}. It seems likely that his awareness that several fundamentally different yet plausible mechanisms for ```1/f" noise already existed would have contributed to his evident lack of enthusiasm for SOC\footnote{In an interview in 1998 with Bernard Sapoval, archived by the Web of Stories project (\url{http://webofstories.com}), he remarked that: ``... criticality goes beyond what I had in mind. In fact, I think perhaps it goes beyond what is necessary. I have not made up my mind on self-organized criticality, because the characteristic of the question of magnets is that there is a parameter like temperature. At a certain critical temperature very special things happen. The characteristic of phenomena like prices or like turbulence, there's no parameter. Therefore to embed a prime [sic] without a parameter in one which has a parameter, and then argue that this parameter somehow arranges to take its own value is presupposing something that is beyond reality. I mean there are no non-critical situations. So I have not made up my mind about the power of this metaphor. The idea that dependence can be global, that variance can be infinite, and in fact that everything that has been taken as finite without any question in physics or in statistics can, in fact be divergent or zero  [...] is something that did not depend upon any broader conjecture about the {\em causes} of these phenomena. It comes out of efforts to {\em describe} them and has been made unavoidable by those efforts." [Our italics]}

Secondly, the concept of criticality (\Sref{BTWcritical}) was invoked to
explain the scaling (\Sref{scaling}) that was seen in nature, drawing
heavily on the established theory
of continuous (\ie second order) phase transitions, but contrasting it with the new
feature of self-organization (see the quote of \citealp{BakChen:1989},
p.~5 below, ``\quotebracketed{T}here is one area of physics \dots the
critical state is self-organized.''). Self organization was  an essential feature of the argument, in order to explain why critical behavior is apparently so \emph{common} in nature. The
traditional notion of criticality placed it firmly at a singular point
in parameter space, which had to be accessed by ultra-fine tuning, such
as careful adjustment of the temperature in a zero-gravity environment
\citep{LipaETAL:1996}.\footnote{In the language of dynamical systems and
the renormalisation group, critical phenomena as observed at phase
transitions \citep{DombGreenLebowitz:1972} are characterized by a fixed
point that is repulsive in several directions and therefore accessible only
from a very narrow basin of attraction.} In contrast, self-organized critical systems would be dynamically \emph{attracted} to a state
where they display scaling, \ie long-range correlations in time and
space dominate and so bring about a new, effective interaction and global
features very different from the microscopic physics: More is indeed different.

As their conclusion, as mentioned above, they proposed ``the SOC
postulate'': 
\begin{citedquote}{BakChen:1989}{p.~5}
The explanation is that open, extended, dissipative dynamical systems
may go automatically to the critical state as long as they are driven
slowly: the critical state is self-organized. We see  fractals as
snapshots of systems operating at the self-organized critical state.
\end{citedquote}
The first sentence refers to features of SOC systems, which have
subsequently been summarized by Jensen as ``slowly driven interaction dominated
threshold [(SDIDT)] systems'' \citep[][p.~126]{JensenBook}.
\citeauthor{BakChen:1989} stressed openness as a required system property because at a stationary state the flux of otherwise
conserved particles
towards dissipative boundaries was perceived early on as a
mechanism by which fluctuations and correlations are communicated
throughout the system:
\begin{citedquote}{BTW88}{p.~368}
Note that Eq. (3.2) conserves $\sum_n z_n$ except at the boundary, so that any
``excess z'' must be transported to the boundary for global relaxation to
occur.
\end{citedquote}
As a result,
``The boundary cannot be scaled out in the limit of large system
sizes as is usually done in statistical physics''
\citep{PaczuskiBassler:2000}.

As well as open boundaries which do not disappear in the large system limit,  \citeauthor{BakChen:1989} also emphasized dissipation, which should be
understood from a dynamical systems perspective. In the presence of
dissipation, dynamical systems explore a greater amount of phase space,
than they would if subject to the constraint of energy conservation. When the 
statement about ``dissipative dynamical systems'' above was written, in \citep{BakChen:1989}, despite the comment on the flux of a conserved quantity, conservation in the sandpile dynamics had
not yet received much attention and the focus still lay with
the apparent lack of conservation when sand grains slip down a hill
thereby reducing potential energy. \citet{HwaKardar:1989a} and
\citet{GrinsteinLeeSachdev:1990} put particle conservation on the map,
the latter demonstrating that conserved dynamics in conjunction with
non-conserved noise (or with conserved noise and spatial anisotropy)
will generically produce scale invariance.

An interesting distinction between BTW's SOC model and classic forward cascade descriptions of fluid turbulence such as Kolmogorov's 1941 model is thus that dissipation in the former takes place at the boundaries and thus on large scales, while in the latter case it is the smallest \citep{ChapmanWatkins:2009}

In \Sref{phenotypegenotype} we will discuss the sufficient and
necessary conditions for SOC. It remains now to comment on the second
sentence in the quote above from Bak and Chen, ``We see fractals as
snapshots of systems operating at the self-organized critical state.''
One could quite legitimately read this as ``we see {\em all}
fractals as \dots'' rather than ``we see {\em such} fractals as \dots''. The former reading is fully in
line with the concise abstract of the paper: ``Fractals in
nature originate from self-organized critical dynamical processes.''
\citep{BakChen:1989} but nonetheless it is unlikely that BTW really believed that all fractals needed SOC to explain them.

While either version of the claim is bold, it is certainly correct that fractal-like structures in time and space are exactly what characterizes critical systems, so that a claim that (some) naturally
occurring fractals  in ``open, extended, dissipative dynamical
systems'' are self-organized, is in fact identical to a claim that open, extended, dissipative dynamical
systems can develop into a
critical state.  

\section{Criticality and minimal stability }
\slabel{BTWcritical}

Considerable confusion has arisen over the years from the several
meanings of the word ``critical'' in the phrase ``Self-Organized Criticality''.  The
word ``critical'' has a very clear   technical meaning in
statistical mechanics and the theory of phase transitions. That this was indeed the
intended meaning in  \citeauthor{BTW87}'s newly coined term ``Self-Organized Criticality'',
 is clear from rereading their \citeyear{BTW87} and \citeyear{BTW88} papers. Unfortunately, in these same
papers the word ``critical'' was occasionally also used in a more
colloquial sense of a threshold. 

In this section  we  clarify and distinguish three
distinct meanings of the word ``critical'': 
\begin{itemize}
\item critical spatiotemporal correlations, such as those seen at phase transitions
\item critical thresholds, and
\item the value of a global (control) parameter at the critical point.
\end{itemize} 
and illustrate them by use of BTW's famous sandpile model.

\subsection{The BTW Model}
\begin{figure}[t] 
\begin{center}
\includegraphics[scale=0.7]{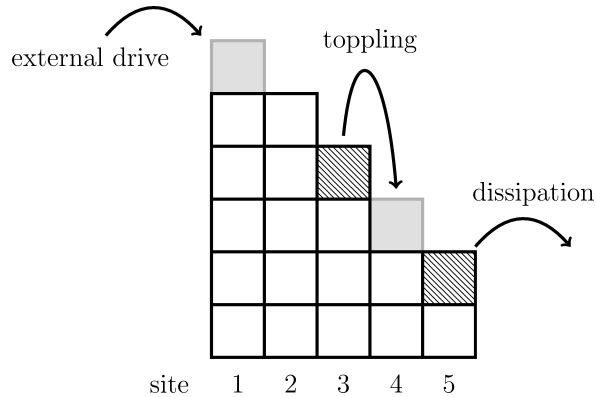}
\end{center}
\caption{\flabel{figure_1dbtw_toppling}The one-dimensional BTW Model on a
lattice of size $L=5$ (for illustrative purposes, this system is
ridiculously small, much bigger systems are normally studied in SOC). Particles which are about to move are shown
hatched, particles which are about to appear somewhere are shown in
gray. The current configuration is $h=\{5,5,4,2,2\}$ and
$h=\{6,5,3,3,1\}$ after the update indicated. One gray particle is
going to be added by the external driving on site $i=1$, which 
takes place only if no toppling occurs somewhere in the systems, such as
the ones on site $5$ or from site $3$ to $4$. The slope
exceeds the threshold at two sites, $i=3$ and $i=5$. When the latter
topples, one particle will be lost by dissipation at the boundary, as if
$h_6=0$ permanently.}
\end{figure} 

In order to fully appreciate the distinction between ``critical'' in the
technical and in the loose sense, it is instructive to introduce the
famous\footnote{Independently invented in the earthquake context by \cite{Katz:1986}} BTW sandpile model, illustrated in 
\Fref{figure_1dbtw_toppling}. On a one-dimensional grid of sites
$i=1,2,\dots,L$ it is defined as follows: Each site $i$ carries a number
$h_i$ of grains. If the slope $h_i-h_{i+1}$ at site $i$ exceeds a threshold,
$h_i-h_{i+1}>1$, then one grain is moved from $i$ to $i+1$, so that
$h_i\to h_i-1$ and $h_{i+1}\to h_{i+1}+1$, thereby decreasing the slope
at $i$ and increasing the slope
at up- and down-stream sites and thus triggering further updates. The
totality of these updates constitutes an avalanche, which is triggered by adding a
particle at a randomly chosen site $i$ (known as the external drive), so that $h_i\to h_i+1$, and
carries on until none of the sites exceeds the threshold any more. Only
then the driving resumes.
In \Fref{figure_1dbtw_toppling} the driving takes place at site $i=1$ 
and a further toppling takes place at site $3$ with $h_3=4$ to site $4$
with $h_4=2$ prior to the update.
The (virtual) edge site $L+1$
carries a stack of height $h_{L+1}=0$ by definition and is never updated,
\ie particles are dissipated here, as indicated in
\Fref{figure_1dbtw_toppling}.

The sandpile model in one dimension has some very distinctive features and
its (very simple) behavior then differs significantly from that seen in dimensions greater than
unity.  In that case, a local, \emph{scalar} slope (rather than a
gradient) is introduced. In two dimensions sites are labeled $(i,j)$
with $i,j\in\{1,\dots,L\}$ and carry a local slope $z_{(i,j)}$. If that
exceeds a threshold, say $3$, then $z_{(i,j)}$ is reduced by $4$, \ie
$z_{(i,j)}\to z_{(i,j)}-4$, and the slope at all four nearest neighbors
increased by one unit, \ie $z_{(i\pm1,j\pm1)}\to z_{(i\pm1,j\pm1)}+1$.
Boundary conditions are such that (virtual) sites outside the lattice are not
updated, \ie slope units are being lost.

The density of these slope units, $\zeta=\sum_{ij} z_{(i,j)}/L^2$ may be
thought of as a control parameter which is generated by the pile rather than externally imposed: If it is big, then avalanches can be
expected to be large --- some sites surely exceed the threshold if
$\zeta$ does. If $\zeta$ is small, avalanches will be small as well.
The external drive increases $\zeta$, at least temporarily, whereas
dissipation at boundaries decreases it. Because large avalanches promote
dissipation, they reduce the control parameter, whereas small ones may
leave it unchanged. In other words, large avalanches can be
expected\footnote{The inversion of this narrative, namely $\zeta$ being
large as a consequence of small avalanches and $\zeta$ being small as a
consequence of large avalanches, is equally convincing --- numerics
suggests further subtleties \citep{PetersPruessner:2009}.} to occur at
comparatively large $\zeta$, typically reducing $\zeta$, whereas small
avalanches occur at low $\zeta$. This feedback has been suggested to be
at the heart of SOC \citep{TangBak:1988b,DickmanVespignaniZapperi:1998}.
Because $\zeta$ is a global average, its fluctuations will decrease with
increasing system size $L$, eventually ``pinching'' it at its mean
\footnote{The reader may be wondering why we did not consider $\zeta$ to be  an order parameter, rather than a control parameter, as it is the response of the system  to the externally controlled drive. If the BTW model had  been a bit more like a conventional critical system  than it has in fact turned out to be, presumably one would also find that the spatiotemporal correlation function: 
\begin{equation}
C(r,t) = <z(r_0,t_0)f(r_0+r,t_0+t)> - <z>^2 
\end{equation}
exhibited algebraic behaviour at criticality. We would then have had the analogy: $\zeta =$ magnetisation, $z =$ local spin, and the external drive $=$ temperature. 

Instead, in our discussion above, and for example also in \citep{PetersPruessner:2009},  $\zeta$ is seen  as the control parameter that drives an activity (topplings). The density of the activity of topplings is then the order parameter of an absorbing state phase transition}.

\subsection{The meaning of ``Criticality''}
\slabel{meaning_of_criticality}
The behavior of the sandpile model is  reminiscent of the
behavior of a system undergoing a continuous phase transition
\citep[\eg][]{Stanley:1971,Yeomans:1991,ChristensenMoloney:2005}. Phase
transitions have been one of the centers of attention in statistical mechanics for
well over one hundred years. They normally occur in a
system as some control parameter, such as the temperature, is changed.


{\bf Critical spatiotemporal correlations:} One particular class of phase transitions, so called continuous or
second-order phase transition, have the peculiar feature that \emph{at
the critical point}, that is for some special value of the control
parameter, correlations become long ranged (follow a power law) and,
equivalently,
fluctuations occur \emph{on all length scales}, \ie
there is no characteristic size and the size distribution of the
fluctuations displays a power law dependence with a non-trivial
exponent. Moreover, an observable indicates the onset of long-range
order, whose presence distinguishes two different phases. Traditionally,
that observable goes by the name of the ``order parameter''. It is a
suitably but not uniquely defined quantity that vanishes in one phase
(the disordered or high-temperature phase) and is finite in the other
(the ordered or low-temperature phase). The susceptibility, which
measures fluctuations and equivalently (by the linear response theorem) the response of
the order parameter to a small external
perturbation, diverges with the system size. This is known as
\emph{critical behavior}, a
\emph{critical phenomenon} or just \emph{criticality}. 

The term self-organized criticality refers to exactly that last, technical
usage of criticality. \citeauthor{BakChen:1989} explicitly referred to
the long range spatial correlations:
\begin{citedquote}{BakChen:1989}{p.~5}
\quotebracketed{T}here is one area of physics where the relation between spatial
and temporal power law behavior is well established. At the critical
point for continuous phase transitions, the correlation function for the
order parameter decays spatially as $r^{2-d-\eta}$ and temporally as
$t^{-d/z}$.\footnote{As the two point correlation functions are normally
expected to decay like $r^{-(d-2+\eta)}\mathcal{G}(r^z/t)$ it is in
order to remark that the temporal dependence would normally expected to be
$t^{-(d-2+\eta)/z}$.} But in order to arrive at the critical point, one
has to fine-tune an \emph{external} control parameter such as the
temperature or the pressure, in contrast to the phenomena above which
occur universally without any fine-tuning. The explanation is that open,
extended, dissipative dynamical systems may go automatically to the
critical state as long as they are driven slowly: the critical state is
self-organized.
\end{citedquote}
Bak, Tang and Wiesenfeld also explain what makes criticality so relevant
and so attractive:
\begin{citedquote}{BTW87}{p.~382}
At the critical point there is a distribution of clusters of all sizes;
local perturbations will therefore propagate
over all length scales, leading to fluctuation lifetimes over all time
scales. A perturbation can lead to anything from a shift of a single
pendulum to an avalanche, depending on where the perturbation is
applied. The lack of a characteristic length leads directly to a lack of
a characteristic time for the resulting fluctuations. 
\end{citedquote}

Criticality in SOC however is not reached by setting a control parameter
to a ``critical'' value:
\begin{citedquote}{BTW87}{p.~381}
The criticality in our theory is fundamentally different from the
critical point at phase transitions in equilibrium statistical mechanics
which can be reached only by tuning of a parameter, for instance the
temperature. The critical point in the dynamical systems studied here is
an attractor reached by starting far from equilibrium: The scaling
properties of the attractor are insensitive to the parameters of the
model. This robustness is essential in our explaining that no fine
tuning is necessary to generate $1/f$ noise (and fractal structures) in
nature.
\end{citedquote}

{\bf Critical thresholds:} Unfortunately a second, looser meaning of ``critical'' has led to confusion. It 
refers to the threshold that is frequently thought to govern the
microscopic dynamics of SOC systems, and can be found in the same
publications as those quoted above.
\begin{citedquote}{BTW87}{p.~382}
This will cause the force on a nearest-neighbor pendulum to exceed the
critical value and the perturbation will propagate by a domino effect
until it hits the end of the array.
\end{citedquote}
Here ``critical value'' refers to the threshold beyond which
activity sets in. 
We would discourage this usage of ``critical'' and instead urge the use
of the alternative, ``threshold value''. We suggest to refrain from
combining it with ``critical'',  as in ``critical threshold''.

{\bf Critical global control parameters:} In a subtle variation of the first meaning of ``critical'' mentioned
above, a third, technical meaning refers to a \emph{global control parameter} taking a critical value:
\begin{citedquote}{BTW87}{p.~382}
If the slope is too large, the pile is far from equilibrium, and the
pile will collapse until the average slope reaches a critical value
where the system is barely stable with respect to small perturbations.
\end{citedquote}
Because it was hitherto unclear whether there were generally order and control
parameters in SOC and, if so, what they were,
one could possibly interpret this quote also as referring to a global \emph{order}
parameter which reaches the value characteristic of criticality.

In traditional critical phenomena, the global control parameter would be
the critical temperature, or a critical probability \etc, but here it is
the \emph{average} slope, which seems to link also to the second
meaning, because it is the average over some local dynamical feature.
In contrast to the first meaning, ``critical'' in the quote   above refers
to the value some parameter attains so that the system maintains
criticality, \eg divergent susceptibility. ``Self-Organized
Criticality'' would then refer to the self-organization of the presumed
control parameter to its critical value, rather than the
self-organization of the system to display criticality. The difference
between the two is obviously subtle, but very important. 
The former interpretation emphasizes the existence of a critical point
and the self-organisation of a control parameter to that value, whereas
the latter focuses on the appearance of the system as critical.
Although we believe that the name SOC refers to the latter (the system
displaying criticality), in \Sref{phenotypegenotype} we briefly discuss  BTW arguments
that a critical point exists and that the system's dynamics drives that
control parameter towards that value.

\subsection{Minimal stability}
 
Further confusion has arisen from the usage of the term ``minimally
stable'', alluding to chaotic behavior which was being explored in
the
literature under the headline ``edge of chaos''
\citep[\eg][]{Langton:1990,KauffmanJohnsen:1991,RayJan:1994,MelbyETAL:2000}.
\begin{citedquote}{BTW87}{p.~384}
Our picture of $1/f$ spectra is that it reflects the dynamics of a self-
organized critical state of minimally stable clusters of all length
scales, which in turn generates fluctuations on all time scales. 
\end{citedquote}
In fact, SOC was introduced using the terminology of ``minimally stable
states'' \citep{TangETAL:1987}, which lose stability by even the tiniest
perturbation anywhere. The language and the basic concept draw on the
theory of dynamical systems.

This seems to be the obvious interpretation of ``minimally stable'',
namely a state where the smallest perturbation leads to a system-wide relaxation. That
is in fact the case in the one-dimensional sandpile model, which
develops into a state where (almost) all sites have a slope
corresponding to the threshold value. However, the one-dimensional
sandpile is exceptional in that respect:
\begin{citedquote}{BTW88}{p.~365}
\quotebracketed{W}e consider
for pedagogical reasons an example in one spatial dimension. In
this case the spatial degrees of freedom ``decouple'' and the system ends
up in the least stable metastable state. This minimally stable state is
a trivial critical state with no spatial patterns and uninteresting
temporal behavior.
\end{citedquote}
In higher dimensions, a small perturbation may lead to a response by the
system at \emph{any} scale. 
In contrast to ``least stable metastable state'', the term ``minimally
stable'' in the context of SOC refers to the \emph{possibility} of system-wide events.
\citet{BTW87} illustrated that very clearly in their Fig.~1, where
every site is shaded that takes part in a particular avalanche. Some
sites among these clusters surely have a slope below the threshold, so
even though labeled ``minimally stable'' the system shown is not in a
state where any charge anywhere would result in a system-wide avalanche
or, in fact, necessarily any avalanche at all.

In fact, later studies make it abundantly clear that the average value
of the dynamical variable (the local degree of freedom subject to interaction, see below) in SOC systems is normally well away from the
threshold. For example, in two dimensions, the sandpile model has been
conjectured \citep[Grassberger quoted by][finally confirmed analytically by
\citealp{CaraccioloSportiello:2012}]{Dhar:2006} to have average height
of $17/8=2.125$, well below the threshold of $3$, and the Abelian Manna Model
\citep{Manna:1991a,Dhar:1999c} with threshold $1$ in one dimension has
average height $0.9488(5)$ \citep{DickmanETAL:2001}, expected to drop to
$1/2$ with increasing dimension \citep{HuynhPruessnerChew:2011}.

In summary, the system's ``critical'' features which \citeauthor{BTW87}
claimed to have been self-organized are   long-ranged correlations and
divergent
susceptibility to external perturbations. SOC systems organize
themselves to a state where they look very much like those at a 
critical point, as if they were undergoing a phase transition. What BTW did not
mean, and did not imply, is that the system organizes itself into a state
where every local degree of freedom is close to some threshold. This
happens to be the case for the one-dimensional sandpile model, but this
should be regarded as a coincidence, not least because the
one-dimensional sandpile model shows no interesting features otherwise.
\citeauthor{BTW87} left it open whether the apparent control parameter
reaches some critical value. In fact, a successful theory of SOC suggests
\citep{DickmanETAL:2001} that the control parameter fluctuates about its
critical value.

\section{The necessary and sufficient conditions for SOC}
\slabel{phenotypegenotype}
The seperation of cause and effect has long been problematic in much of the debate surrounding SOC, so we now set out  arguments for,
\begin{itemize}
\item three necessary features that a system needs to exhibit in order to qualify as SOC
\item and three sufficient ingredients comprising a mechanism  (SDIDT) that produces SOC
\end{itemize}
and we attempt to decide if SOC and SDIDT are synonymous.

\subsection{SOC's ``phenotype": The necessary conditions to observe it}

As explained above, in \Sref{BTWcritical}, BTW regarded SOC as a
critical phenomenon in the traditional sense of statistical mechanics, 
\ie a system displaying non-trivial scaling (scaling that deviates from
what is generated by simple dimensional analysis, see \Sref{scaling}). While
this can be seen in many different observables, and in fact, in SOC is
often observed in integrated, global quantities, such as avalanche sizes
and durations, scaling \emph{should} manifest itself in particular through the presence of
long-ranged spatio-temporal correlations. The term ``long-ranged''
alludes to the fact that, again, these correlations should display
power law scaling and not decay like, say, an exponential. Demanding
direct evidence for the scaling of spatio-temporal correlations is a
technical challenge, as explained in \Sref{scaling}. 

One key aspect of SOC, however, deviates strongly from traditional (tuned)
critical phenomena (see the quote above in \Sref{meaning_of_criticality}, 
``\quotebracketed{T}here is one area
\dots continuous phase transitions
\dots without any fine-tuning
\dots critical state is self-organized.'' from \citep[][p.~5]{BakChen:1989}), in that these always require tuning to a critical
point, \ie a precise setting of one or more parameters to a specific
\emph{finite} value.\footnote{The finiteness (neither vanishing nor infinite) is
crucial, because any finite value provides a \emph{scale} which other
characteristic scales can compete with, and which allow the formation of
dimensionless ratios of quantities, that can be raised to arbitrary
powers. Only in the presence of such competing length-scales can one
expect to see non-trivial scaling exponents, see Appendix A.} 

The dynamics of SOC systems is supposed to drive them to the critical
point without the need of such external ``tweaking''  of a control
parameter. Prior to the advent of SOC, some systems were known, in particular
growth phenomena \citep[\eg the KPZ equation, ][]{KardarParisiZhang:1986}, that did display
non-trivial scaling without external tuning of a control parameter (and
certainly in the presence of competing length scales). BTW, however, 
argued that although a critical point exists in SOC systems, the
\emph{dynamics itself} drives the system towards a critical point,
which otherwise would only be reached by external tuning of a control
parameter:
\begin{citedquote}{BTW87}{p.~381}
The critical point in the dynamical systems studied here is an attractor
reached by starting far from equilibrium: The scaling properties of the
attractor are insensitive to the parameters of the model. This
robustness is essential in our explaining that no fine tuning is
necessary to generate $1/f$ noise (and fractal structures) in nature.
\quotebracketed{\dots}
In a sense, the dynamically selected configuration is similar to the
critical point at a percolation transition where the structure stops
carrying current over infinite distances, or at a second-order phase
transition where the magnetization clusters stop communicating.
\end{citedquote}

A suitable mechanism of self-organization to a critical point, in some ways reminiscent of earlier
suggestions by \citet{TangBak:1988a,TangBak:1988b}, was put forward and
made explicit by \citet[][also
\citealp{DickmanVespignaniZapperi:1998}]{VespignaniETAL:1998}. Its
verification remains subject to ongoing research.

Having made the case for ``truly self-organized, truly critical'' SOC
systems, it remains to remark that even so, every finite system still has an
inherent scale, namely  the system size itself. This is obviously also the case
in traditional critical phenomena, but there the control parameter can
be tuned away from the critical point. In such tuned systems the phenomena
observed and the measurements taken can therefore approximate the infinite system
(or ``thermodynamic limit''), at least in principle, arbitrarily well, by
increasing the system size ever more, as the control parameter is tuned
closer and closer to the critical value. This
is not the case in SOC systems, which are (supposed to be) located right
\emph{at} the critical point with the result that all observables that
display any form of scaling, or which are expected to be divergent in the
thermodynamic limit, will depend on the system size. This dependence is
called finite size scaling, a well known and understood  aspect of  traditional critical phenomena
\citep[][see also the discussion in Appendix A]{Barber:1983}.

The key-features of a system in the SOC state, their \emph{phenotype},
can thus be summarized as
\begin{enumerate}
\item Non-trivial scaling (finite size scaling; no dependence on a
control parameter).
\item Spatio-temporal power law correlations.
\item Apparent self tuning to the critical point (of a possibly
identified, underlying continuous
order phase transition).
\end{enumerate}
where the first and the second item may be seen as aspects of the same
feature: criticality.

Apart from the many proposed SOC systems,
one candidate-system that seems to fulfill (some of) the above features
is invasion percolation \citep{WilkinsonWillemsen:1983} which predated
SOC and was  likened to it early on \citep{GrassbergerManna:1990}. It clearly
displays non-trivial scaling (namely that of percolation), it clearly
displays spatial correlations, and (with suitable definition of the
dynamics) also correlations in time.  In fact, much like traditional SOC
models, its burst-like evolution may be regarded as one possible form of 
intermittency - an aspect of many complex systems, to be further discussed below. However, such avalanches do not
appear in cycles of charge and relaxation, but are part of an ever
increasing region invaded by a cluster
\citep{SornetteJohansenDornic:1995}. In other words, invasion
percolation does not develop into a stationary state (in the statistical
sense)\footnote{Invasion percolation is nonstationary in the same way that an ordinary Brownian random walker is nonstationary.  It keeps growing with time, similar to the way in which the random walker keeps wandering further and further away from $x=0$, leading to $ \langle x \rangle^2>\propto t$.}. What is more, even when invasion percolation displays the
scaling of ordinary percolation, there is no suggestion of any self-tuning taking
place. Rather, invasion percolation sits right at the critical point by
definition. In that sense, invasion percolation resembles Brownian motion,
which offers a rich variety of power law correlated features (although
not strictly non-trivial), without any apparent self-tuning \citep{Sornette:2006,Milovanov:2013}.

\subsection{SOC's ``genotype": The sufficient conditions to generate it}

Above we have relied heavily on the original, early work by BTW to
\emph{define} SOC
phenomenologically. The features
listed above are in fact all the \emph{necessary} conditions for SOC --- with ``necessary" taken strictly in the logical sense (they define SOC: if and only if all of them are observed, one is faced with
SOC). The \emph{sufficient} conditions for SOC then point to a
\emph{cause} of SOC, asking for the system's key ingredients in order to
produce those SOC characteristics, and are in a sense its genotype. 
A lot of the research into SOC centers precisely around these sufficient
conditions; the early hunt for different members of the BTW universality class
\citep[\eg][]{Zhang:1989,Manna:1991a} was clearly motivated by this
question. 

The most obvious key-ingredient of any SOC model is the presence of
non-linearities in the interaction, so that the response is not a
simple, linear function of the size of the external perturbation.
\footnote{Without non-linearity, trivial exponents follow. In that sense,
demanding non-triviality implies non-linearity of an interaction which
dominates the global behavior of the system.} In most SOC \emph{models}
the non-linearity is realized as a threshold in the interaction, \ie
activity can spread only
when some local dynamical variable exceeds a
threshold.  We note that the algorithms for BTW's sandpile model and the Edwards-Wilkinson (EW,\citep{EdwardsWilkinson:1982}) model of deposition are the same except that BTW SOC has thresholded diffusion whereas EW has simple diffusion.

When the threshold is overcome, interactions between neighboring
dynamical variables take place (often referred to as ``topplings'') and
as a result bursts of activity occur in the
form of avalanches, which can involve the system in its entirety. These
avalanches spread because the interaction in a toppling can induce a
neighboring local dynamical
variable to overcome  a threshold, even when prior to the interaction it
was not very close to being ``triggered''. 
In the presence of thresholds, avalanching is thus naturally expected.
Of the sufficient conditions listed below, avalanching is therefore the
most likely candidate to be superfluous, because it is implied by the
other conditions, and in fact may be listed with the necessary ones, in
the sense that it is part of the definition of SOC beyond the immediate
meaning of just these three letters.

Surveying the wealth of systems (supposedly) displaying SOC, 
a very strong candidate for our final key-ingredient is the separation of the  time
scales of driving and relaxation, which is implied already in the original definition of the BTW
model \citep{BTW87}. SOC systems are \emph{slowly driven}, so that the
characteristic time scale of the driving does not interfere with
the internal, fast time scale of the relaxation. In computer models and
generally when the relaxation occurs in bursts or avalanches, the
separation of time scales can be completed.\footnote{Yet, in some systems a
second separation is necessary
\citep{DrosselSchwabl:1992a,ClarDrosselSchwabl:1996} which amounts to a
careful adjustment to parameters \citep{JensenPruessner:2002a,BonachelaMunoz:2009}}
 
If one insists on intermittent relaxation in the form of avalanches,
then it is obvious that the driving must be sufficiently slow as not
to disturb the avalanche while it is running \citep{ChapmanWatkins:2009,ChapmanEA:2009}, otherwise continuous activity will result \citep{CorralPaczuski:1999} and individual avalanches are
no longer discernible without the use of, say, some arbitrary threshold.

In particular in the earlier days of SOC, some debate
\citep[][]{DickmanVespignaniZapperi:1998,BrokerGrassberger:1999} evolved
around the question whether the demand
of a 
separation of time scales amounts to a form of tuning or global
supervision \citep[by a ``babysitter'', \citealp{DickmanETAL:2000}, or a
``farmer''][]{BrokerGrassberger:1999}, thereby rendering
SOC a tuned type of criticality after all. 

We thus summarize the \emph{sufficient} conditions so far (the ``\emph{genotype}'', to draw that parallel) as
\begin{enumerate}
\setcounter{enumi}{3}
\item Non-linear interaction (required by 1), normally in the form of
thresholds.
\item Avalanching (intermittency, expected in the presence of
thresholds and slow driving).
\item Separation of time scales (obvious requirement to sustain distinct
avalanches).
\end{enumerate}

We hypothesise that every system that simultaneously fulfils these three conditions will display SOC according to its definition in 1-3, and vice versa.

Together with the ``phenotypical'' conditions 1--3 listed above one might summarize all six of them by
defining SOC as ``Slowly driven, avalanching (intermittent) systems with
non-linear interactions, that display non-trivial power law correlations
(cutoff by the system size) as known from ordinary critical phenomena,
but with internal, self-organized, rather than external tuning of a
control parameter (to a non-trivial value).''

\subsection{Must SOC and SDIDT always be the same ?}

We believe that the most promising instances of SOC, in particular
computer and theoretical models, fulfill these criteria. If they are
really sufficient (but see below) and 
implied by the necessary conditions,\footnote{It is
a strong claim that a system that organizes itself to a critical point
is necessarily intermittent.}
\ie they are complete and not too narrow, then
fulfilling conditions 1--3 implies
fulfilling conditions 4--6 and vice versa. 

\begin{figure}[t]
\begin{center}
\subfigure[SDIDT is a subset of IDT and of
SOC.]{\includegraphics[height=1.9cm]{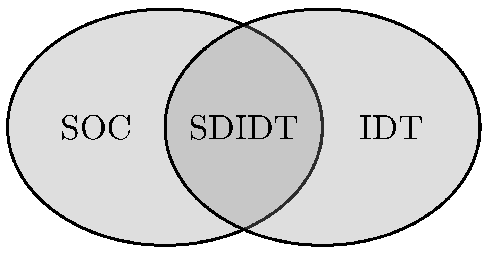}}
\ \ \subfigure[Maybe all of SOC belongs to
SDIDT.]{\includegraphics[height=1.9cm]{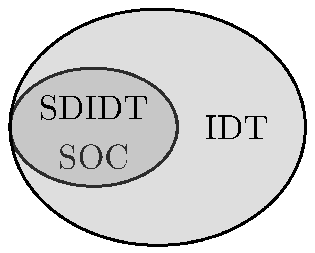}}
\ \ \subfigure[Maybe SOC is a small subset of
SDIDT.]{\includegraphics[height=1.9cm]{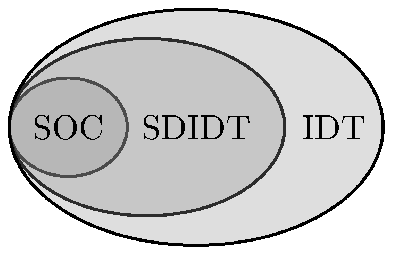}}
\end{center}
\caption{\flabel{venn}
Slowly driven, interaction dominated threshold (SDIDT) systems are a
subset of interaction dominated threshold (IDT) systems. Interpreting slow
drive as implying intermittency, SDIDT coincides (roughly) with the list
of sufficient conditions, 4--6, for SOC. All SDIDT may therefore be expected
to display SOC, as indicated by both Figure 1.3(a) and
Figure 1.3(b). In the latter Venn diagram, however, \emph{all} SOC
belongs to SDIDT, \ie SDIDT and therefore condition 4--6 are not only
sufficient, they are also necessary. It may well be, however, that
conditions 4--6 are not complete, \ie they are only a subset of the
sufficient conditions. In that case, Figure 1.3(c)
applies, where SOC is a subset of the larger class of SDIDT.}
\end{figure}

Conditions 4--6 may also be interpreted as a paraphrase of ``slowly
driven, interaction dominated threshold (SDIDT) systems'' (see
\Sref{BTWpostulate}), with 
intermittency implied by the slow drive. That SDIDT is
sufficient for the occurrence of SOC was conjectured earlier
\citep[][p.~126]{JensenBook} and was well received in the plasma
science community, \Sref{SOCwild}, although sometimes with much reduced emphasis
on slow drive. Such systems are shown as IDT in the Venn diagrams \Fref{venn}. There are several models
considered for their supposed SOC behavior, which either lack any
obvious driving or whose driving is subject to conditions beyond just
being slow
\citep{WilkinsonWillemsen:1983,MossnerDrosselSchwabl:1992,JensenPruessner:2002a,JensenPruessner:2004,BonachelaMunoz:2009}.
To this day, SOC models are
studied with finite drive \citep{CorralPaczuski:1999}, \ie as IDT models in
their own right. The Venn diagrams \Fref{venn} exclude the possibility
of all IDT being SOC,
as the only overlap of SOC and IDT amounts to SDIDT, indicating that
slow drive may not only be a sufficient condition for SOC, but also a
necessary one, which sits well with the notion that finite driving
introduces a finite scale and therefore possibly a cutoff.

While the relation between SDIDT and IDT is an obvious one, the relation between
SDIDT and SOC is less clear. According to the list above, points 4--6, SDIDT
systems should display SOC (as indicated in Fig 1.3(a) 
and 1.3(b))
but that SOC phenomena are restricted to SDIDT is a
much stronger statement, as illustrated in Fig 1.3(b).
 If we
identify criteria 4--6 with SDIDT and they are not too narrow (in that
sense truly minimal), then there
is no SOC outside SDIDT, \ie Fig 1.3(b)
and not Fig 1.3(a) 
is the correct representation of the \latin{status quo} of SOC.

Given, however, that some supposed SOC models, like the Forest Fire Model lack
scaling \citep{PruessnerJensen:2002}, a more realistic concern is that
conditions 4--6 are incomplete, so only some particular SDIDT systems
display SOC, as shown in Fig 1.3(c).
It remains one of the most
important questions in SOC to complete the list of sufficient conditions
without making them too narrow.

\section{Why then is SOC controversial?} 
\slabel{controversy}
In response to the points we have made above, a natural question 
may already be occurring to the reader: ``If, as you claim,  SOC was originally relatively
clearly defined, and if one can now define necessary and sufficient conditions for it, why was it (and is still) controversial''? This is a good question, and there are many reasons, of which we identify  the following   as particularly important: 
\begin{itemize}
 \item Uncertainty and miscommunication about what the essential SOC
 claim in fact was; confusion between the phenomena to be explained and
 the mechanism proposed as their explanation; and, as a consequence,
 confusion about what to look for as experimental ``proof", and what to look for as
 potential application of the theory.
 \item SOC models and supposed occurrence of SOC in nature are easy to
 test for badly and difficult to test for well. To this day, outside the field of tuned phase transitions, really solid
 empirical evidence for scale invariance in nature by lab experiment or
 analysis of observational data is actually quite limited even when its ``ubiquity'', as recognized by Mandelbrot and Bak, is the
 very motivation for the field.  Debates thus continue over the ubiquity of
 fractals, particularly spatio-temporal fractals and avalanches
 \citep[\eg][]{AvnirETAL:1998}.
 \item Many of the SOC models are highly idealized and do not even
 attempt to capture the basic interactions of a natural system. Rather than the caricature of magnetism in the Ising model, or the way in which a shell model encapsulates the symmetries of turbulence, they often
 consist of a set of rules in the vein of a cellular automaton, designed
 to display spatio-temporal scale invariance. However, at closer
 inspection many fail to display the desired features. 
 \item ``Human Factors'': Citation and priority of Mandelbrot and others ruffled some feathers, while SOC may have been a distraction from other important and prior work on spatio-temporal fractality.
 \item Confusion about the deterministic nature and predictability in
 some SOC models, and the natural phenomena they were supposed to apply
 to. On the one hand, the prime example of SOC was the sandpile model
 which evolves according to deterministic rules, on the other hand,
 \citet[][p.~15636]{BakTang:1989a} concluded (about earthquakes) that ``there is virtually
 no hope for ever making specific predictions''.
 \item There are alternative explanations even within other theoretical work on  critical phenomena for ``dirty power laws'' and ``fat tails'', such as ``plain old criticality''
 \citep{PerkovicDahmenSethna:1995} or ``sweeping of an instability''
 \citep{Sornette:1994}.
\end{itemize}

In the following, we address these points in further details.

\newcommand{\paraheader}[1]{\noindent\textbf{#1}\newline\noindent}
\renewcommand{\paraheader}[1]{\subsection{#1}}
\paraheader{Confusion}
\slabel{confusion}
As discussed in \Sref{WhichSoc} and Figure 1.1, the relatively clear core claim of
SOC (of the possibility of self-tuned phase transitions in nature) was sometimes coupled with a perception that SOC
aims to explain \emph{all} fractals or even \emph{all} power laws.
That claim was not made initially, though some proponents of SOC
later nourished that belief even in their popular writing. In Bak's own ``How
Nature Works'' \citep{Bak:1996} power laws of possible relevance to SOC
(such as solar flare X rays) and others almost certainly irrelevant to
SOC (such as Zipf's law of word length) were mentioned side by side. This
had the unfortunate consequence of obscuring the fact that the power law \emph{distributions} in
avalanche sizes and durations were only proxies for the power law
\emph{correlation functions} that BTW described as a crucial aspect of
the unification of spatio-temporal fractals that they were seeking
\citep[][also \Sref{BTWmotivation} and
\Sref{phenotypegenotype}]{BakTangWiesenfeld:1988}. 

This already serious problem was
compounded by a confusion of the proposed explanation (SOC and its
dynamics) with the explanandum (the thing to be explained, namely [ubiquitous] spatio-temporal
fractals). For as soon as
the dynamics of SOC processes is accepted as the universal explanation for the
phenomenon of spatio-temporal fractals, every observation of such
fractals becomes an instance of ``SOC at work''. Worse,
observation of scaling may be seen as \emph{evidence} for the
validity of SOC as an explanation, and (by association or a leap of faith) of the boldest of all SOC claims (\Fref{soc_diagram}), that the contingency of nature derives from SOC.

At first, measuring event size distributions may have been a necessary
evil, as correlations are so much harder to acquire and analyze (see
Appendix A), at
least in
numerical simulations and from observational data.
Over the years, in the absence of an easy way of measuring 
correlations, the literature as a whole moved towards the notion of
power law distributions as a \emph{replacement} (rather than a
proxy, as discussed in \Sref{BTWmotivation}) for power law correlations. To a large extent the distinction was
forgotten and the significance of the latter associated with
the former. Statistical mechanics provides a systematic link
between the two in the form of a sum-rule, akin to the one relating
susceptibility and correlation functions in, say, the Ising Model
\citep[][p.~120, and in SOC, \citealp{PruessnerBook}, Sec.~8.5.4.1]{Stanley:1971}. 
Yet, the relation between the two is strained
by technicalities and it is often far from obvious which correlation
function is expected to display (power law) scaling if an observable
representing a spatio-temporal integral (such as the avalanche size as
the activity integrated in time and space) follows a power law
distribution. Power law distributions are therefore often a proxy for
something unknown.

Nevertheless, a significant
number of papers in the wider literature accepted the notion that every
observation of a power law readily signals the presence of long ranged
(\ie power law)
spatio-temporal correlations. In some cases, power law distributions are
``trivial'' in that they arise without non-trivial interaction and
correlations (see Appendix A).
For example, some directed sandpile models display power law scaling in
the avalanche size distribution, but no spatial correlations whatsoever
\citep{Pruessner_exactTAOM:2003}.
It is fair to
assume that the proponents of SOC were well aware of the difference
between ``power law distributions'' on the one hand and
``power law correlations'' on the other. It is probably also fair to assume that they were
fully aware of the core claim of SOC being an hypothesis subject to an
ongoing investigation.

Above we have tried to draw a line between power laws observed in event
size distributions and power laws observed in correlations. Not every power law event size
distribution is indicative of power law correlations. 
Traditionally, at least in statistical mechanics, the emphasis has been
on the latter, as power law correlations indicate long ranged
correlations, which normally (if exponents are not too large) signal
cooperative phenomena. They are interesting, because the whole is then more
than its parts, \ie the system cannot then be completely described by decomposing it
into smaller compartments or components. 

The distinction
between ``spatio-temporal (power law) correlations'' and
``spatio-temporal fractals'' is even more blurred: Clearly, fractal
spatio-temporal structures imply
non-trivial, long-ranged (\ie power law) spatio-temporal correlations.
The converse connection, however, is quite loose, as it
is far from clear as to \emph{what}
to expect to be fractal in the presence of power law correlations. Is it
justified to assume that
fractal features of less tangible objects (such as spatio-temporal
activity patterns) indicate an underlying fractal structure of the
constituent parts of the system?

\paraheader{Ubiquity, universality, generality}
The argument ``\dots it is hard to believe that long-range spatial and
temporal correlations can exist independently \dots \quotebracketed{a}nd
a large, coherent spatial structure cannot disappear (or be created)
instantly.'' \citep[][p.~5, as quoted in
\Sref{BTWpostulate}]{BakChen:1989}, is a reasonable one and won many
supporters. It rests on the realization that surely long-range
correlations cannot be confined to a particular dimension, rather they
feed through to \emph{all} space and time dimensions. That perception,
however, has long been revised: In equilibrium critical phenomena,
certain dynamics or algorithms \citep{SwendsenWang:1987,Wolff:1989}
evolve spatial fractals essentially with little or no temporal
correlations. 
Vice versa, temporal correlations do
not necessitate spatial correlations, as illustrated by, say, directed
sandpile models \citep{DharRamaswamy:1989,Pruessner_exactTAOM:2003},
which carry no spatial correlations and yet display memory. In fact,
\citet[][p.~267]{Grinstein:1995} called temporal correlations
``in the presence of a local conservation law \quotebracketed{\dots} difficult to avoid''.
In other words, if
SOC is expected in the simultaneous presence of spatial and temporal
correlations, then the existence of \emph{both} has to be ascertained,
because one does not imply the other.

As mentioned several times throughout this piece, it is notoriously difficult to measure long range spatio-temporal
correlations \latin{in situ} or even numerically (but see below) and
therefore many authors resorted to measuring \emph{spatio-temporal
integrals} of observables, such as avalanches sizes, durations and
areas. The scaling of the distribution of these event sizes, say the
avalanche size, can be related to the scaling of a correlation function,
say the activity propagator, measuring the spreading of ``activity'' in
the system some time after a
triggering event at some ``seeding point'' in the system
\citep{Pruessner:2015:FT}. Alternatively, the avalanche size can be
expressed in terms of 
the spatially averaged activity
\citep[][Sec.~9.3.4 and \citealp{McAteerETAL:2014}]{Luebeck:2004,PruessnerBook}.
Less directly, the width of the
interfacial mapping \citep{PaczuskiBoettcher:1996,Pruessner:2003} of the Oslo Model 
\citep{ChristensenETAL:1996}, which is related to the scaling of the
probability density function of the avalanche size,
scales exactly like the height-height correlation function of that
interface \citep{BarabasiStanley:1995}. It is probably fair to say, that
it is difficult enough to extract from
experiments and observations \emph{any} scaling or fractality, which are therefore seen as ``good
enough'' substitute or, more accurately, symptoms of the long range spatio-temporal correlation
supposedly causing them.

Where fractals and scaling are suspected in natural phenomena,
observational support is often very limited \citep[\eg]{AvnirETAL:1998}, both in terms
of length and time scales spanned by the data as well as its robustness.
Broad distributions are frequently found, but there are few
phenomena, which offer sufficiently detailed and broad data to support power law
scaling beyond reasonable doubt. 
It is difficult to reconcile the
efforts that have been spent on experiments, data gathering and analysis
with the claim that scaling or just power laws are ubiquitous in nature. One
may therefore ask, rather provocatively: Is there really a (ubiquitous)
problem to solve?

Unless one accepts the claim that SOC is the basis of scaling in
nature, SOC itself (not just scaling) as defined in \Sref{phenotypegenotype} is difficult to
identify in a natural phenomenon or experiment directly. If anything, SOC has
been offered as an explanation for certain scaling to
appear \emph{spontaneously}. At the theoretical end, none
of even the \emph{computer models} which are widely accepted as
displaying all the hallmarks of SOC (see \Sref{paradigmatic_vs_good}) has been solved or even only
systematically approximated. In fact, there is not even a mean field
theory that makes any quantitative reference to SOC taking place in
spatially extended systems with some form of boundary at finite
distances. 

In summary,
SOC was conceived as an explanation of a ubiquitous natural phenomenon,
but it turns out that observational or experimental evidence is
very difficult to come by. Hard evidence for SOC is mostly due to numerical
modeling. To this day, there is no complete theory of SOC and it
remains unclear why a phenomenon, that should be observable under 
generic conditions is so rarely seen. 

In that particular respect, SOC  has shared the
fate of the ``directed percolation universality class''
\citep[\eg][]{Hinrichsen:2000a,Odor:2004}, which, although widely accepted to apply
to an enormously large class of phenomena
\citep{Janssen:1981,Grassberger:1982}, ranging from catalytic chemical reactions or to epidemic spreading, still has very little
experimental and observational support
\citep[][however see the laboratory experiments beginning with  \citealp{TakeuchiETAL:2007} and accompanying news coverage e.g.  \citealp{Hinrichsen:2010},
and the intriguing observational claim of   \citealp{WanlissUritsky:2010}]{Hinrichsen:2000b}.
 
\paraheader{Paradigmatic versus good models}\slabel{paradigmatic_vs_good}
SOC has been introduced and motivated by the sandpile model, which is
given in the form of a set of updating rules as used for the description
of cellular automata. The initial numerical analysis revealed what was
then coined ``Self-Organized Criticality'' and $1/f$ noise, later
revised to be $1/f^2$ by \citet[][also
\citealp{ChristensenFogedbyJensen:1991}]{JensenChristensenFogedby:1989}. The model itself was
early on revised to display the Abelian property \citep{Dhar:1990a}, which is beneficial to
both numerical and theoretical analysis. Over the years, it
became increasingly clear that the sandpile model has some rather
unfortunate features, in particular, that its supposed scaling behavior
could never be fully determined
\citep[\eg][]{Manna:1990,LuebeckUsadel:1997a,DeMenechStellaTebaldi:1998,DornHughesChristensen:2001}; The prime model of Self Organised Criticality turns out 
not to display much of that notorious Criticality after all. On the other hand, it offers a
vast array of secondary features that had very interesting large scale
properties which have been characterized analytically, such as waves
\citep[\eg][]{IvashkevichKtitarevPriezzhev:1994a},
the average slope \citep[\eg][]{JengPirouxRuelle:2006}, (static) height-height correlation function
\citep[\eg][]{Jeng:2005b} or solvable variants with anisotropy
\citep{DharRamaswamy:1989}.
None of this work, unfortunately, makes reference to scaling of avalanches, large scale activity correlations or spatio-temporal
fractals, although the sandpile model certainly carries similar visual
appeal \citep{Creutz:2004}.

As far as ``real sandpiles'' are concerned, experimental studies failed
to detect robust scaling \citep{JaegerLiuNagel:1989,HeldETAL:1990},
although, as one may argue, expecting otherwise would stretch the name  ``sandpile
model" beyond its intention as aide-memoire. One should remember that even the first SOC papers discussed a coupled harmonic oscillator model as well as the sandpile. Interestingly, the ricepile
experiment \citep{FretteETAL:1996} and the ricepile or Oslo model
\citep{ChristensenETAL:1996} both fared much better in that respect. 
As far as granular media is concerned, the Oslo model  has probably the
best
experimental support
\citep{FretteETAL:1996,AhlgrenETAL:2002,AegerterGuntherWijngaarden:2003,LorinczWijngaarden:2007}.

The Oslo Model is, in fact, a 
representative of an entire universality class
\citep{NakanishiSneppen:1997}, often referred to as the Manna
universality class. Equally, the Manna Model \citep{Manna:1991a,Dhar:1999a} displays most clearly
all features one could
possibly expect from a self-organized critical model
(\Sref{phenotypegenotype}): 
\begin{itemize}
\item Firstly, robust,
reproducible finite size scaling without dependence on any control
parameter or details of the definition of the model
\citep{DickmanTomedeOliveira:2002}, such as the
underlying lattice structure
\citep{HuynhPruessnerChew:2011}. 
\item Secondly,
spatio-temporal correlations, which were initially measured through
integrated observables (avalanche size, duration, area, radius of
gyration \etc). While temporal correlations are less of a concern
\citep[\eg][for correlations on the slow time
scale]{PickeringPruessnerChristensen:2012}, spatial correlations can be
extracted with some patience \citep{McAteerETAL:2014}. 
\item Thirdly, apparent
self-tuning to a critical point, that can be characterised in its own
right, \ie as a regular critical point without invoking SOC
\citep{VespignaniETAL:1998,DickmanETAL:2001}. 
\end{itemize}
In fact, it seems that two important theoretical tools are within reach for 
the Manna Model: an  $\epsilon$-expansion  
\citep{HuynhPruessner:2012b}, and a
field-theoretic description which also reveals the universality class of
a tuned variant \citep[the conserved directed percolation universality
class according to][]{RossiPastor-SatorrasVespignani:2000}.
The Manna Model also fits the list of ``ingredients'' of an SOC Model in
\Sref{phenotypegenotype}: Thresholds, intermittency and separation of
time scales. The universality class of the Manna Model is remarkably
large \citep[][p.~177--181]{PruessnerBook}, containing even fully
deterministic models \citep{deSousaVieira:1992,PaczuskiBoettcher:1996}.

Going back to $1/f$-noise as the motivation and root of SOC,
\citet{Jensen:1990} introduced a fully deterministic lattice gas inspired by experimentally
observed $1/f$ spectra in superconductors. Simulations of this model
exhibit $1/f$ spectra and the dissipation take place on fractal-like structures. However, recently it was realized that the model does
not display self-organization to criticality
\citep{GiomettoJensen:2012}, but requires tuning to reach the critical
point of the (conserved directed percolation) absorbing state phase transition.
It is probably fair to say that despite its long history
\citep{vanderZiel:1950} $1/f$-noise is no longer a motivation for
SOC, possibly because of the confusion about its actual meaning ($1/f$
versus $1/f^{\alpha}$) and also the possibility, at least in contemporary
computer models, to characterize correlations
directly in the time domain rather than indirectly via the power spectrum.

\paraheader{Distraction and priority}
Some of the early papers in SOC paid insufficient attention to, and so may have led other
people to neglect, related (and previous) relevant work. Bak and Chen
openly declared that they could see little collaboration between those
working on fractals and those working on $1/f$ noise \citep[``\quotebracketed{\dots} those working on fractal phenomena \quotebracketed{\dots} never \quotebracketed{\dots}
seem to be interested in the temporal aspects, \quotebracketed{\dots} those working on ``1/ f'' noise never bother with the spatial structure of the source of the signal''][as quoted in
\Sref{BTWmotivation}]{BakChen:1989}. Yet, laboratory critical phenomena
already linked space and time, for example via critical slowing down,
which is \emph{exactly} the concept used to understand dynamical
critical behaviour (e.g. \cite{Yeomans:1991}) in SOC. There was also work on the link between
spatial and temporal fractality by Mandelbrot himself \citep{MandelbrotWallis:1969} prior to
his work on spatio-temporal cascades in turbulence beginning in the 1970s
\citep{Mandelbrot:1972}. 

In the early days, some scientists may have perceived SOC as an
aggressive foray into their established scientific fields, an attempted
``hostile takeover'', which contributed to the notion of ``physicist
hubris'' \citep[also][]{Maddox:1994}. The Bak-Sneppen Model, for example, was introduced to the biologist audience  by summarizing their own achievements and contrasting them with those of the
authors: ``However, there is no theory deriving the consequences of
Darwin's principles for macroevolution. This is a challenge to
which we are responding'' \citep[][p.~5209]{SneppenETAL:1995}. Plenty of similar examples can be found in
the literature, some witty, some outright rude \citep[``Is biology
too difficult for biologists?''][]{Bak:1998}. His fellow  complexity scientists  Cosma Shalizi and Bill Tozier at the Santa Fe Institute penned an amusing riposte to this tone in their preprint ``a simple model of the
evolution of simple models of evolution'' \citep{ShaliziTozier:1999}.

\paraheader{Predictability} 
Predictability has a somewhat ambiguous status in SOC. In their second paper,
BTW argued for $1/f$ to be the result of a  superposition
of independent avalanche durations \citep[][p.~369]{BTW88}, as originally
suggested by \citet{vanderZiel:1950}. In other words, independence of events
was an assumption at the very foundation of SOC as an explanation of $1/f$
noise. 
Although convenient for a straight-forward quantitative relationship
between $1/f^{\alpha}$ exponent and avalanche duration distribution, however,
independence is not needed for the argument about the origin of $1/f$ noise.
Once introduced, the implied
lack of predictability and generally contingency became an important
feature, a ``selling point'', very early, for example in the work on
earthquakes mentioned above, but also as part of the wider perspective
of SOC:
\begin{citedquote}{BakPaczuski:1995}{p.~6690}
The \quotebracketed{SOC} system exhibits punctuated equilibrium behavior, where
periods of stasis are interrupted
by intermittent bursts of activity. Since these systems are noisy,
the actual events cannot be predicted; however, the statistical
distribution of these events is predictable. Thus, if the tape of
history were to be rerun, with slightly different random noise,
the resulting outcome would be completely different. Some
large catastrophic events would be avoided, but others would
inevitably occur. No ``quick-fix'' solution can stabilize the
system and prevent fluctuations. If this picture is correct for
the real world, then we must accept fluctuations and change as
inevitable.
\end{citedquote}
Although the authors stress here that it is the noise that is inherently
unpredictable, its input is ``amplified'' even in fully deterministic SOC
systems, because of their high susceptibility to external perturbation,
which is characteristic for chaotic systems (\eg the famous ``butterfly-effect'') and those at a critical point:
\begin{citedquote}{BakTang:1989a}{p.~15636}
At several points the earthquake is almost dying, and its
continued evolution depends on minor details of the crust of the earth far
from the place of origin. Thus in order to predict the size of the
earthquake, one must have extremely detailed knowledge on very minor
features of the earth far from the place where the earthquake
originated.
\end{citedquote}

On the other hand, long temporal correlations, both at the fast
time scale within an avalanche and the slow time scale between
avalanches, mean that the system maintains a memory of past activity.
In systems where a globally conserved quantity is released in sudden
bursts, this is immediately obvious: A slow external drive will
eventually ``run out of steam'' to sustain large events in quick
succession. In these cases, one can expect anti-correlations
\citep{WelinderPruessnerChristensen:2007}. In general, long
temporal correlations allow for particularly good predictability, at
least of big events. This does not contradict the notion of
large susceptibility, which indicates that the \emph{variance} of
responses to an external perturbation is particularly broad. Clearly,
all of these observables are probabilistic by nature.

In a Nature debate on earthquake prediction, \citet{Bak:1999} later
qualified and clarified his views on predictability:
\begin{citedquote}{Bak:1999}{}
\quotebracketed{T}he earthquakes in SOC models are clustered in time and
space, and therefore also reproduce the observation O4
\quotebracketed{seismicity is not Poissonian}.
This implies that the longer you have waited since the last event of a
given size, the longer you still have to wait; as noted in Main's
opening piece, but in sharp contrast to popular belief!

\quotebracketed{\dots}
For the longest time-scales this implies that in regions where there
have been no large earthquakes for thousands or millions of years, we
can expect to wait thousands or millions of years before we are going to
see another one. We can 'predict' that it is relatively safe to stay in
a region with little recent historical activity, as everyone knows.
There is no characteristic timescale where the probability starts
increasing, as would be the case if we were dealing with a periodic
phenomenon.

\quotebracketed{\dots}
Unfortunately, the size of an individual earthquake is contingent upon
minor variations of the actual configuration of the crust of the Earth,
as discussed in Main's introduction. Thus, any precursor state of a
large event is essentially identical to a precursor state of a small
event. The earthquake does not ``know how large it will become'', as
eloquently stated by Scholz. Thus, if the crust of the earth is in a SOC
state, there is a bleak future for individual earthquake prediction. On
the other hand, the consequences of the spatio-temporal correlation
function for time-dependent hazard calculations have so far not been
fully exploited!
\end{citedquote}
In the same piece, Bak also acknowledged
(and rejected) a differing perception of SOC:
\begin{citedquote}{Bak:1999}{}
The \quotebracketed{SOC} phenomenon is fractal in space and time, ranging from minutes and
hours to millions of years in time, and from meters to thousands of
kilometers in space. This behaviour could hardly be more different from
Christopher Scholz's description that ``SOC refers to a global
state\dots containing many earthquake generating faults with uncorrelated
states'' and that in the SOC state ``earthquakes of any size can occur
randomly anywhere at any time''.
\end{citedquote}
 
It seems that the understanding of predictability in SOC became more
differentiated over time. While initially the insight prevailed that
stochasticity and susceptibility made SOC systems inherently
unpredictable, that view made way for a better understanding of
correlations. SOC systems do not signal the onset of a large event and
may not even do so while the event occurs. Yet, event sizes remain
correlated over very long time, allowing \emph{probabilistic}
predictions, such as the likelihood of two particularly large events
occurring consecutively.

In SOC-inspired research in solar physics, waiting time distributions (WTDs) are the most prominent
format of predictions. They are defined as the probability density
function of the waiting times between consecutive events. A
Poisson process produces an exponentially decaying waiting time
distribution \citep{vanKampen:1992}, which is therefore often used as the
fingerprint for a lack of correlations. However, non-stationary point
processes may give rise to (apparent) power law tails in the WTD
\citep[][Ch.~5]{Aschwanden:2011}. In the literature WTDs based on
observations of solar flares have given varying results depending on the
observational period, the X-ray wavelength, and whether individual
active regions are considered in the analysis. \citet{CrosbyETAL:1998}
found no correlation between the elapsed time interval between
successive deka-keV solar flares arising from the same active region,
and the peak intensity of the flare. This observation was taken to be
\emph{in support} of the solar flare SOC model by
\citet{LuHamilton:1991}. In contrast, based on soft X-ray flare
observations, \citet{BoffettaETAL:1999} found that the WTD displayed
power law behavior in contradiction with the SOC model by
\citeauthor{LuHamilton:1991}, which predicts Poisson-like statistics.

To put this apparent mismatch in perspective, we want to emphasize that
Poissonian waiting times, or more generally, lack of correlations are by
no means typical in SOC. For example, the Omori-law of earthquakes
\citep{Omori:1894,Utsu:1961,UtsuOgataMatsuura:1995} plays a very prominent
r{\^o}le \citep{OlamiChristensen:1992,HergartenNeugebauer:2002} in the analysis of the SOC model
by \citet{OlamiFederChristensen:1992}. 
One can only speculate whether the presence of Poissonian waiting times,
$\mathcal{P}(t)=\lambda \exp(-\lambda t)$ for a process with rate $\lambda$,
may have been confused with a power law distribution of waiting times,
in the limit of small $\lambda$ (namely with large cutoff), as $\lambda
\exp(-\lambda t)=t^{-1}\mathcal{G}(\lambda t)$ with $\mathcal{G}(x)=x\exp(-x)$, the scaling
function.

\paraheader{Alternative scenarios}
There have also been a number of successful attempts to provide 
alternative explanations for (apparent) critical behavior without
tuning of a control-parameter. \citet{Sornette:2006} has collected a
number of scenarios under which apparently critical behavior can be
observed without invoking SOC. One described very early
\citep{Sornette:1994} suggests that an ordinary critical phenomenon is
causing the scaling behavior, yet no self-organization takes place
beyond the system's tendency to remain close to the critical point: The
system ``sweeps back and forth'' across the critical point in an oscillatory fashion.

\citet{PetersNeelin:2006} performed an analysis reminiscent of one
done when dealing with equilibrium continuous phase transitions. They
studied precipitation of rain by identifying the water vapor density as
the control parameter (in analogy with the temperature in a ferromagnetic phase transition) and identified
the amount of precipitation as the order parameter. In addition they
plotted the variance of the precipitation, and also how frequently the
atmosphere is found at a given vapor density, which they call the
residence time. The outcome is a set of diagrams which exhibit many
similarities to how the order parameter and susceptibility behave in
standard continuous phase transitions. The precipitation picks up
abruptly at a certain vapor density and in the vicinity of this density
they find that the fluctuations in the precipitation (corresponding to the susceptibility)
peaks.

However, as the atmosphere
does not self-tune to a particular critical value of the vapor density,
but rather is found in a range of vapor densities.  The near critical
behavior is related to the residence time having a peak near the value
of the vapor density at which the precipitation has a sharp increase and
the fluctuations in the precipitation peaks.

This analysis may be interpreted in the following way. The dynamics of
the precipitation pulls the atmosphere around the critical value of
the vapor density as vapor may build up beyond the critical value
and rain showers can take the vapor density back down below the
critical value. As a result the atmospheric systems  moves
around a certain vapor density at which precipitation becomes very
likely but sharp tuning to a critical state does not take place. 

A very similar analysis and scenario was found for the activity of the
brain in resting state as measured by fMRI by
\cite{Tagliazucchi:EtAl2012}. These authors analyzed the brain activity
from the perspective of percolation and found that the brain moves
around in the vicinity of a three dimensional percolation transition for
the voxel activity measured by an fMRI scanner. 

These results may suggest that at least in some situations the SOC
phenomenology in reality consists of dynamics that by itself drives the
system to the neighborhood of some critical transition but, which,
because of coupling between the dynamics and the order parameter, is
unable to fine tune to the exact critical state. In an attempt to
provide a theoretical foundation of SOC, it has been argued 
\citep[][but \citealp{PetersPruessner:2006}]{VespignaniETAL:1998,DickmanVespignaniZapperi:1998} that this is
a matter of system size: As the system size is increased, the dynamics
is eventually ``pinched'' at the critical point. In the case of 
precipitation it appears that the order parameter (amount of rain) is
able to pull the control parameter (vapor density) below the critical
value and that the control parameter (due to the build up of super
critical vapor densities by evaporation) can grow above the critical
value. This seems to be similar to the dynamical cause
\citep{PruessnerJensen:2002} that breaks the scaling of the
\cite{DrosselSchwabl:1992a} forest fire model.

\section{SOC in the wild: how has SOC  inspired research on space and fusion  plasmas?} 
\slabel{SOCwild}

The previous section may read like a catalogue of woe, and it is important to see things in perspective. A theory as bold as SOC was bound to be controversial, so we will now balance the controversy with a very brief sketch of  how the research fields of three of the authors (Chapman, Crosby and Watkins), in solar system and laboratory fusion plasmas, have been inspired by the SOC paradigm into new and productive directions. We direct the reader in search of more detail to the companion papers in this volume \citep{AschwandenETAL:2014,McAteerETAL:2014,SharmaETAL:2015}, the reviews of \cite{Chapman:2001,Watkins:2001,Watkins:2002,Freeman:2002,Vassiliadis:2006,DendyChapmanPaczuski:2007,PerroneETAL:2013}, the book by \cite{Aschwanden:2011}, and references therein,  among many possible sources.
 
Several problems in space plasma physics resemble SOC. One clear example
is the wideband distribution of solar flare energies, and solar flares
remain one of the most intriguing examples of SOC-like behavior.    Most
likely caused by a magnetic instability that triggers a magnetic
reconnection process in a large range of sizes and time scales, solar
flares produce emission in almost all wavelengths (e.g. gamma-rays, hard
X-rays, soft X-rays, extreme ultraviolet , Hydrogen $\alpha$ emission,
radio wavelengths, and sometimes even in white light). 
\cite{DatloweElcanHudson:1974,LinETAL:1984} and \cite{Dennis:1985}  were some of the first to
determine frequency distributions of solar flare hard X-ray observations
(see \cite{CrosbyAschwandenDennis:1993} for a historical summary).  

From micro- and nano-flares to the largest flares the flare energy power
law distribution is found to cover over eight orders of magnitude
\citep{Aschwanden:2011}. The energy distribution contains
all flare sizes, independently of the mechanism by which the released
energy is converted. Like the Gutenberg-Richter law in seismicity these
observations predated SOC, and \cite{LuHamilton:1991} proposed a model
based on BTW's sandpile to reproduce them. In their model each solar
flare is considered an avalanche event in a critical system. The way the
magnetic energy is redistributed, how the system is driven (the
``loading mechanism''), and the ``incorporation'' of
magnetohydrodynamics (MHD) have all been further developed by others, and interesting SOC-inspired variants have also been proposed such as the cascade of reconnecting loops studied by   \cite{HughesETAL:2003}.

Following   \cite{LuHamilton:1991} several workers  began analyzing
frequency distributions  on large solar flare datasets in the context of
SOC (e.g. \cite{CrosbyAschwandenDennis:1993,  CrosbyETAL:1998,LeePetrosianMcTiernan:1993,GeorgoulisVilmerCrosby:2001}. Many studies also followed that used solar flare
observations in other wavelengths.   \cite{CrosbyETAL:1998}, for example, subdivided
solar flare X-ray data according to a parameter and determined
frequency distributions on the resulting sub-sets,
revealing positive correlations in the parameters. In the context
of model validation, observational results such as these put constraints
on models that need to be able to reproduce the observations.

Turning now to the Earth's local plasma environment, the magnetosphere, our readers, whether space scientists or not, will know of the dramatic auroral displays seen over Earth's polar regions. These  reveal a range of intricate patterns, and  many phenomena have been identified in them on a wide range of temporal and spatial scales, from seconds to hours, and from one to thousands of kilometers (e.g. panels A to C of the figure in \cite{Freeman:2002}). In the early 1990s some researchers began to focus on whether there might be ``universal" aspects to auroral structure. As well as chaotic nonlinear dynamics, SOC was a natural avenue for this inquiry, and several parallel lines of attack developed. We will mention just a few  papers here, a more comprehensive bibliography of early work on magnetospheric SOC can be found in \cite{Watkins:2001}. 

One strand was experimental.   \cite{TakaloEA:1993} for example computed structure functions (as also widely used in turbulence research and surface growth) on the auroral electrojet (AE) index\footnote{AE is a time series of the peak magnetic perturbation measured on the ground caused by electrical currents flowing 100 km overhead in the aurora}.  They  found a scaling region between about 1 minute and 2 hours. The scale break above 2 hours was attributed to the  quasi-periodic interruption of the time series by a global scale auroral disturbance, the magnetospheric substorm.

A complementary theoretical thread took several forms. The strand  most directly inspired by sandpile models initially involved pointing out the similarities between key properties of AE, determined by power spectral (e.g. \cite{Consolini:1997,UritskyPudovkin:1998}) and threshold exceedence (e.g. \cite{Consolini:1997}) techniques, and those of existing sandpile models, both BTW's and the running sandpile model of \cite{HwaKardar:1989a}. The pioneering work on power spectra of AE by \cite{Tsurutani:1990} that showed it to exhibit a low frequency ``1/f" region, was now argued to be indicative of SOC. Two new measurements directly inspired by SOC  were the probability density of the time for which the AE index exceeded any given fixed threshold, the ``burst duration", and burst size (the integrated value above the threshold for each burst). Both were found to have fat tailed pdf's (e.g. \citep{Consolini:1997}),  for bursts from the minimum measurement scale of 1 min to the longest burst lifetimes (of order 1 day).  Subsequent work (e.g. \citep{FreemanETAL:2000}) for both burst duration  found that superposed on the fat tail was another component centered on a fixed scale of about 100 min, corresponding to the global substorm phenomenon.

In parallel with the above developments in space physics, SOC  had also already  been fruitful in fusion research, where the wider properties and dynamics of avalanching systems  are of interest in addition to their statistical properties. It had been noted ( e.g.  \cite{DendyChapmanPaczuski:2007}) that magnetically confined tokamak plasma experiments for fusion are driven, dissipative systems with multiple steady states, anomalous transport, and bursty release of energy and material. This prompted the development and extensive study of several sandpile/avalanche models (surveyed in \cite{DendyHelander:1997,PerroneETAL:2013}) in the fusion context, specifically to reproduce key observables which are not necessarily power law avalanche distributions. A key observable in tokamaks is the correlation between the  distinct statistical properties of bursty energy release (edge localised modes) and the confinement state of the plasma (low and high confinement states, or L and H modes).   This essential property was captured in a "sandpile with an H mode" \citep{Chapman:2000,ChapmanETAL:2001}. This model also captures aspects of anomalous transport in tokamaks, for example the observed, unexpected, inward transport against the temperature gradient. This fusion-relevant model directly followed from  one developed \citep{Chapman:1998} to explore the role of SOC in magnetospheric substorms, and is an example of transfer of ideas from one research area to another and back.  The CDH model  \citep{Chapman:1998} which could be consistent both with  fat tailed ionospheric energy dissipation events, and with magnetospheric  events with a characteristic size provided that they were systemwide events like the substorm, was inspired by  work on inertial sandpiles for tokamaks \citep{DendyHelander:1998},   

Following \cite{Chapman:1998} a more direct observational test was suggested in  \cite{LuiETAL:2000}, who proposed the use of a  threshold exceedance measure to investigate the spatial structure of the aurora.  Using UV images of the aurora from cameras on the Polar spacecraft, \cite{LuiETAL:2000} identified auroral "blobs", where the auroral emission intensity exceeded some fixed threshold, during both quiet and substorm intervals.  As in the AE index time series analysis,  a fat tailed pdf was found both for number of threshold exceedances and their areas, with an additional population centred on a fixed scale   corresponding to the global substorm disturbance. It was also realised that unlike the ideal SOC paradigm, that in such driven dissipative astrophysical confinement systems the driving would be highly variable, leading to studies of the extent to which the fat tailed avalanche distribution was robust against this \citep{Watkins:1999}.  

However, \cite{Uritsky:2002}  argued that the \cite{LuiETAL:2000} approach overestimated the number of spatio-temporally evolving blobs, because a blob counted in one image at one time could be the same one counted at another time. These authors thus analysed Polar  images from  spatiotemporally, and claimed that pdfs of maximum blob area or integrated area over blob lifetime followed  power law distributions over the entire observable range (3-5 orders of magnitude), and similarly for the blob lifetime, maximum dissipated power and dissipated energy (see also \cite{Freeman:2002}).   

    The conceptual parallel between avalanches in SOC models with those apparently observed in the aurora is appealing, but an immediate complication resulted from the fact the aurora is a projection of the dynamic charged particle structure of the near-Earth magnetosphere.  Because  satellite measurements in the tail  region of the magnetosphere have shown bursty bulk flows of charged particles that may be individually correlated with auroral emissions, and may have a scale-free distribution of durations, it was argued \cite{LuiETAL:2000,Uritsky:2002} that these were, essentially, the avalanches.  

A somewhat different, and complementary scenario to BTW's SOC for the dynamic structure of the magnetosphere was however suggested by \cite{Chang:1992}.  In his picture, plasma wave resonances create coherent structures of various sizes that merge and interact to create new structures.  He proposed that continual interactions of this type may naturally self-organize or be forced into a scale-free hierarchy of coherent structures like the ordering of spin structures in the Ising model at the critical point.  In his view the distinction between self-organized and forced criticality is essentially about the nature of the thing that drives the system (the solar wind in the case of the magnetosphere).  As we have shown, in BTW's SOC, the driving rate is necessarily very slow compared to the interaction and merging time scales.  Intriguingly the opposite behaviour  was predicted for some other non-equilibrium models where the onset of criticality appears above some driving rate \citep{NicolisMansour:1984}, and something analogous is also seen in turbulence where the onset of complex behaviour occurs {\em above} a given value of the Reynolds number. \cite{ChapmanWatkins:2009,ChapmanEA:2009} have clarified this behaviour by noticing that the dimensionless control parameter formed by fuelling rate and dissipation rate in SOC models is effectively an inverse Reynolds number.

Consideration of the driver has however, as elsewhere in complexity research, raised a thorny issue: The supply of energy from the solar wind into the magnetosphere has itself a fractal flavour, because the solar wind is turbulent. Studies \citep{FreemanETAL:2000} using static measures of fractal property in long-of order years- non-overlapping solar wind and auroral time series suggested that, at least for the AE index, the scale free behaviour might originate in the solar wind, rather than be self-organised in the magnetotail. Comparisons using time-dependent measures on shorter-of order months or less-but overlapping, series however \citep{UritskyETAL:2001} indicated that an internally generated scale-free component may coexist with a solar wind. Debate on this topic has continued, and is not unique to the magnetosphere, but is reminiscent, for example of the debate in theories of punctuated evolution between the influence of  ``external" events (such as asteroid impact) on extinctions and self-organised ``internal" extinctions.

Even without an SOC origin, power law distributions can be used to estimate   the maximum
strength of natural hazards  and are increasingly being used by reinsurance
companies and governments to assess the risks they pose.  The space industry is no exception to this trend, as when building spacecraft  such information is essential when designing
the spacecraft shielding which mitigates against extreme events
as well as the long-term effects  of space weather. 

\section{Summary and conclusion}
\slabel{conclusion}

Readers who have made it to the end of this article may now appreciate why our first epigraph quoted the Dude from ``The Big Lebowski", as untangling the history, meaning, and current status of SOC really has required the reader (and authors) to keep track of a ``lotta strands". This is made even harder by the diversity of the research fields in which these strands originate, all of which have not only their own notations and traditions, but also very different ideas about what a good model is, and how to wield Occam's razor !  However, we hope we have also brought out the reasons why our second epigraph quoted physics Nobelist Philip Anderson,  who described SOC as of ``paradigmatic value, as the kind of generalization which will characterize the next stage of physics". In our concluding section we now try to draw out two specific issues, about the current status of the SOC conjecture and accompanying theory and  the testability of SOC in space and lab plasmas respectively, and give our views on these.

\subsection{SOC theory: where do we stand ?}
SOC was conceived by Bak, Tang and Wiesenfeld against the
background of condensed matter theory, statistical mechanics and, to
lesser extent, dynamical systems, with the intention to explain
spatio-temporal fractals in nature. 
The initial core claim,   that some spatio-temporal fractals (\ie long time
and range correlations) are produced by systems that are organizing
themselves to a continuous phase transition, where such correlations
are typical, was soon extended to encompass a much
greater spectrum of phenomena. Considerable confusion has grown over the
years as to what has been established by and about SOC, to what extent
it has been confirmed analytically, numerically, by observation in
nature or experimentally, where it applies and what it aims to explain.

There are few systems that display SOC in all its glory, but they {\em do}
exist and they provide clear evidence that it works in precisely the way
originally envisaged. SOC may be at work in some natural phenomena, such
as earthquakes, solar flares and precipitation, but SOC is almost certainly not
ubiquitous. To some, more traditionally-minded communities, in
particular in condensed matter theory, the phenomenon of SOC
nevertheless comes as a great surprise, as spontaneous non-trivial
scaling in this area is otherwise confined to systems displaying generic
scale invariance, without intermittency or self-organization to a critical
point, and invariably requiring some scale-free source or input, such as
noise.

Despite being hampered by re-interpretations not originally intended by its
authors (and sometimes because of these !), SOC has inspired much research into multiscale phenomena and
has helped bring together disjoint communities, in particular those
interested in heavy tails, spatio-temporal fractals and $1/f$ noise. All
of these were known to specialists
\citep[\eg][]{vanderZiel:1950,SchickVerveen:1974,Weissman:1988}.
but all had relatively low cross-disciplinary visibility before SOC, as the authors can testify. In
the long term this may be one of the most important legacies of the
subject.

While, in some of these areas, the strict definition of SOC has given
way to a broader view and sometimes sweeping claims, it has also
provided the very fruitful paradigm for a much deeper understanding of
the phenomena concerned, as researchers became aware of the distinct
possibility that some very simple interactions on a microscopic scale
carry over to and evolve across many different time and length scales,
effectively providing the same basic physics in rescaled form across
many scales. In that sense, SOC realized the aspirations and exhortations of
\citet{Anderson:1972} and \citet{Wilson:1979} in that it provided a
framework to ask questions about the crucial, effective, simple
interactions that are present across all scales of a multiscale
phenomenon, and which must therefore be present, detectable and
describable (in bare, unscaled form) at some small scale. SOC stripped
away the need for a detailed microscopic physics and gave way to a more
global perspective of the \emph{basic physical principles} that govern a
phenomenon on every scale. 

This perspective of looking for the basic interaction that governs a
physical system across scales is different from classic reductionism,
which suggests that the overall phenomenon is some averaged version of
the internal dynamics. SOC suggests that the interaction is
present on all scales, although in some scaled form, as it slowly
morphs and evolves in space and time. In that respect, SOC provides a
much sharper quantitative emphasis than some of the more recent
complexity-inspired points of view.

Broad, heavy tailed distributions and correlations, whether or not they can
justifiably be called power laws, and regardless of whether they are
indicative of scaling, are observed in many field and pose a challenge.
This is because they suggest that phenomena are not confined to a particular
length scale and that the physics driving them manages to cross scales.
To understand them better would allow a better quantitative
characterization of fluctuations and associated risks and is likely to
point at the relevant underlying physics. SOC is one such attempt at a
better understanding. With all its flaws and shortcomings, it is
difficult to identify a more successful approach.

\subsection{Testing for SOC in space plasmas} 
Having clarified our best current understanding of what an SOC state is, the other key question is then, are these solar flare, and other interesting
plasma systems, really in this SOC state? Testing for SOC has mainly been
centered on testing for power law statistics of event sizes.
Observationally this presents a fundamental challenge, as the confined
plasma systems are of finite size. The solar corona offers the broadest
range of spatial scales and indeed, here we see power laws over up to 8
decades. Probability distributions of different auroral ``spot" variables (observed in earth's ionosphere), constructed using the results of ground-based and satellite camera observations, extend across more than two orders of magnitude in space \citep{KozelovUritskyKlimas:2004}, enabling the derived burst variables (which convolve a  time variable) such as size, duration and so forth to span a much larger range . This distinction between spatial and spatiotemporal scaling ranges has been well known for some time, see e.g. \cite{AvnirETAL:1998}.
A second challenge is that the developing understanding of
how to precisely test for SOC, as discussed above, has ``raised the
bar'' in terms of what is required for a truly convincing demonstration.
Solid data of spatio-temporal correlations remain out of reach and thus
global measures, such as spatial (activity) integrals have to be used.
We  will only touch on some points here, see also \citep{McAteerETAL:2014}

First, it is important to distinguish SOC, or indeed, multiscale
avalanching, from turbulence, and this follows from the intrinsic
separation of timescales in these systems. The idealized SOC limit is
when the ratio of driving and dissipation is taken arbitrarily small, and
this is in the opposite sense to turbulence \citep{ChapmanWatkins:2009,ChapmanEA:2009}. The finite
size of these systems makes distinguishing SOC and turbulence in these
plasmas from observations of the scaling properties alone a challenge
and this has led to controversy \citep{UritskyETAL:2007,WatkinsChapmanRosenberg:2009}. 

Second, one must exclude ``trivial'' similarity that can occur in linear
systems. A simple Brownian walk is self-similar but does not imply
spatio-temporal correlations. An example of a spatially extended system
that incorporates
dynamics is the Edwards-Wilkinson model \citep{EdwardsWilkinson:1982}, where grains are randomly
dropped onto a surface which is smoothed by linear spatial diffusion. In
such a model one observes power laws in the sizes of patches of the
surface where the height exceeds a threshold, however the model is
linear and in that sense trivial  \citep[][also Appendix A]{ChapmanETAL:2004}. However, even
escaping triviality and linearity, MHD plasmas, along with
hydrodynamics, exhibit similarity in their non-linear dynamics, yet are
certainly not instances of SOC. Classic
examples are non-linear Alfven waves, shocks and solitons. It therefore
does not suffice to look for non-trivial power law event size statistics per se as
the ``hallmark'' of SOC. As discussed in section 1.7 it  is necessary, but not sufficient.

Given the difficulties inherent in observational verification of
(idealized) self-similarity, alongside the clear evidence for multiscale
bursty energy release and the simultaneous operation of a zoo of plasma
processes operating on multiple spatio-temporal scales which are strongly
coupled to each other, the original SOC paradigm can be said to have
``mutated'' into a broader concept of ``multiscale avalanching'' plasma
systems. As a concept around which to order the
observations, multiscale avalanching has been a remarkable success. 
Without the concepts of plasmas as multiscale systems (e.g. Chang, 1992),
phenomenologists would still be restricted to the detailed plasma physics of
an energy release event in isolation. Avalanching, that is, bursty transport and energy release events on multiple
scales, is observed to be  ubiquitous in driven, dissipative plasmas, and involves  fully non-linear physics,
coupling across multiple scales.  It remains an open, and highly topical problem across astrophysical and laboratory plasmas.

\section*{Appendix A: Dimensional analysis, scaling and self-similarity}
\slabel{scaling}
Scaling is a continuous symmetry obeyed by certain physical observables.
It relates, quantitatively, the value of a physical observable at one set of
parameters to its value at another set of parameters. It comes in two forms: The
trivial form is obtained by a dimensional analysis, the non-trivial
form is the manifestation of self-similarity, identified most
prominently by the renormalization group, but also accessible
numerically and by data analysis (such as a data collapse).

Scaling is a very powerful concept in physics, because it allows the
analysis of a phenomenon on a vastly different scale than it is observed
on. What is more, the \emph{same fundamental physics} must be at work at very
different scales, which often leads to very deep insights. The fact, for
example, that the electrical force between two charges decays like $1/r^2$
carries the signature of the dimensionality of the space around us,
$d-1=2$, and is explained within the framework of Quantum Electrodynamics by the
\emph{masslessness} of the photon.

Scaling can be applied amazingly broadly, as demonstrated in
\posscitet{Buckingham:1914} $\Pi$ theorem which introduced the method of
dimensional analysis more than 100 years ago. The scaling
determined by dimensional analysis is often referred to as trivial: It is an
unavoidable consequence of finding, imposing or assuming a certain
physical reality. For example, \emph{assuming} that the frequency $\omega$ of a
frictionless mathematical pendulum depends only on its length $\ell$,
its mass $m$ and the gravitational acceleration $g$ has the immediate
consequence that it must necessarily be a constant multiple of $\sqrt{g/\ell}$. 

In the small angle approximation the constant is unity --- but even
without the small angle approximation and thus allowing for dependence
on the amplitude $\phi_0$, dimensional analysis tells us that
the frequency \emph{must} be of the form $\omega=f(\phi_0)\sqrt{g/\ell}$, where
$f(\phi_0)$ is an
(\latin{a priori} unknown) function of $\phi_0$. By dimensional
analysis, the observable $\omega$ obeys the remarkable symmetry 
$$
\omega(\phi_0, \ell, g) = 
T^{-1} 
\omega(\phi_0, L^{-1}\ell, L^{-1}T^2g) 
$$
for all finite, positive, real $T$ and $L$. In particular, by choosing
$T=\sqrt{\ell/g}$ and $L=\ell$,
$$
\omega(\phi_0, \ell, g) = 
\sqrt{\frac{g}{\ell}}
\omega(\phi_0, 1, 1) 
$$
and we can identify
$f(\phi_0)=\omega(\phi_0, 1, 1)$.

Dimensional analysis can only ever give rise to trivial 
exponents,
usually integers or simple fractions, which are a necessary consequence
of the dimension of the quantities used to describe the physical
phenomenon and the assumption that the physical reality of a phenomenon
is independent from the choice of units used to describe
it.\footnote{Expressing distances in units of time is an everyday
example, ``It's four hours to Washington.'' \emph{vs} ``It's $260$ miles
to Washington.'' \citep{Pruessner:2004}.} We notice that
``trivial'' is a loaded word, but it is a technical term, used to point
to the fact that the scaling obtained is identical to that found in a
system without considering the effect of non-linearities (which otherwise make ``all the music''). In such a
\emph{linear} system, solutions can be superimposed, suggesting a lack
of interaction. One such solution may be the \emph{trivial solution},
and by association the linear system and its exponents are therefore
called trivial. The term ``trivial'' obscures the fact that there are
famous examples of dimensional analysis producing powerful and
far-reaching results, such as \posscitet{Kolmogorov:1941a} $5/3$ law.
Yet, the deep insight does not consist in the dimensional analysis, but
in determining the physical quantities that enter into a physical
phenomenon. Dimensional analysis is only a relatively straight-forward
manifestation of that achievement.

Trivial scaling does not produce the richness and
variety of power laws found in nature. For example, the fractal
dimension of percolating clusters on a square lattice is $91/48$
\citep{StaufferAharony:1994}. This is generally possible in the presence of
dimensionless, finite ratios involving the characteristic length or
distance (or, more generally, time) the
system is studied under. In case of the pendulum mentioned above, for
example, the initial condition might be expressed as an initial
displacement $d$ of the pendulum, so that $\phi_0=\sin^{-1}(d/\ell)$. In
that case, it is no longer obvious how $\omega$ scales in $\ell$,
everything is possible, at least in principle.\footnote{To claim that
the frequency of a pendulum depends on the absolute value of its initial
displacement suggests a physics different from the one where the
frequency depends only on the initial \emph{angle}.}

The presence of non-trivial power law spatio-temporal correlations indicates on the
one hand that competing scales are present (otherwise exponents are
determined by dimensional analysis), on the other hand that they do not
\emph{dominate} the behavior of the system, in the sense that their
physics does not take over on large spatio-temporal scales.
Rather, they compete with and balance each other. Otherwise a
\emph{characteristic scale} appears, with one ``physics'' below and one
``physics''
above that scale.\footnote{In principle, scaling may be found
nevertheless, as happens in finite systems, which exhibit power law
correlations on a scale small compared to the system size, beyond which
all correlations necessarily vanish.}
A characteristic scale is present, for example when
spatio-temporal correlations decay (asymptotically) exponentially, say
$C(r)=C_0
\exp(-r/\xi)$ with some characteristic scale $\xi$ and unknown amplitude
$C_0$. The system ``knows''
the scale and it can be determined from within, for example as the
inverse slope of the plot of
$\log\left(C(r)/C(2r)\right)=r/\xi$ against $r$. 

When correlations decay like a power law, no such manipulation is
possible, say $C(r)=C_0 (r/\xi)^{-\mu}$ with some unknown exponent
$\mu$. Without knowing $C_0$ the length $\xi$ cannot be extracted.
For example $C(r)/C(2r)=2^{\mu}$. In the presence of power law
correlations, there is a lack of scale from within, in other words, the
system is \emph{self-similar}. Self-similarity generally manifests
itself in the non-trivial scaling of correlation functions.

Power law correlations are typically observed at transitions,
also in dynamical systems, when a fixed point changes stability. They
have been extensively studied within the field of \emph{critical phenomena} ever since
\posscitet{Onsager:1944} solution of the two-dimensional Ising Model
suggested a clash with Landau-theory \citep[][]{Stanley:1971}, which produces the same
exponents as dimensional analysis suggests. Power law correlations are
generally observed at \emph{continuous} transitions, a smooth change of phase
as opposed to the sharp, so-called \emph{first order} phase transitions
as seen when water boils in a kettle. They can be observed in 
the spectacular display of critical opalescence in
carbon dioxide \citep{Stanley:1971}.

As pointed out several times above, correlation functions are difficult to measure directly, in particular
in SOC models which often have open boundaries and are therefore not
translationally invariant. As a result, correlation functions depend
not only on the distance, but on absolute coordinates, so that spatial
averaging is not possible. Moreover, correlations are often very weak, so
that the indirect, integrated measures mentioned above, such as the avalanche sizes and
durations, often show clearer signs of scaling.

The exponents characterizing the power laws are usually expected to be
universal, \ie they are the same in vastly different systems. Because
they are characteristics of asymptotics, which in turn a determined by
basic features of the interactions, exponents are intricately linked to
the \emph{symmetries} of the interactions and the system as a whole. It
was one of the great insights of the renormalization group
\citep{Wilson:1971a,Wilson:1979} that exponents are characteristics of the
symmetries involved. Results are particularly strong for phase
transitions in two-dimensional equilibrium systems with discrete
symmetries: Conformal field theory
\citep{LanglandsPichetPouliotStAubin:1992,Cardy:1992} and
Stochastic Loewner Evolution
\citep{LawlerSchrammWerner:2001,SmirnovWerner:2001} were able to
demonstrate that there are exactly $6$ universality classes 
\citep{Fogedby:2009}.

Exponents are not the only universal quantities. Also universal are
amplitude ratios, moment ratios and scaling functions, although in the case of finite size
scaling they often depend on boundary conditions
\citep{Barber:1983,PrivmanHohenbergAharony:1991}. 
Traditional critical phenomena consider scaling of an observable in an
infinite system as a function of some control
parameter, say the magnetization density $m$ as a function of the 
temperature difference to the critical value $T-T_c$, observing
$m\propto(T_c-T)^\beta$ for $T<T_c$.
Alternatively, a system may be tuned to the critical point and the
scaling of a correlation function is studied as a function of the
distance. At the critical point, system wide observables can be studied
for their dependence on the system size, known as finite size scaling,
say $m\propto L^{-\beta/\nu}$, for a system with linear extent $L$. In
both cases, a length scale (such as the distance or the size of the
system) controls the scaling of the observable in a system right
\emph{at} the critical point. Leaving correlation functions aside,
finite size scaling of some global observables is the only scaling
displayed by SOC.
For example, the cutoff $s_c$ in the
avalanche size distribution scales like $s_c\propto L^D$ as a function of 
$L$. The exponent $D$, the fractal dimension
of the \emph{characteristic} avalanche size, is expected to be
universal.

While exponents are normally not
independent, because they are related by scaling relations, other
quantities are, yet the power of universality ties them up in a
universality class. Only a few universality classes are expected to
exist, so determining a
very small number of universal quantities determines the whole class. 

That is what BTW were envisaging for SOC; universality justifies the
study of toy models:
\begin{citedquote}{BTW88}{p.~365}
We choose the simplest possible models rather than wholly realistic and
therefore complex models of actual physical systems.
Besides our expectation that the overall qualitative features are captured in this way, it is certainly possible that quantitative properties
(such as scaling exponents) may apply to more realistic situations,
since the system operates at a critical point where universality may
apply. The philosophy is analogous to that of equilibrium statistical
physics where results are based on Ising models (and Heisenberg models, etc.) which have only the symmetry in common with real systems. Our
``Ising models'' are discrete cellular automata, which are much simpler
to study than continuous partial differential equations.
\end{citedquote}

The early literature responded to that call for universality by offering
models in the ``BTW universality class'' \citep{Zhang:1989,Manna:1991a}.
The focus quickly shifted to introducing new universality classes, in
particular by breaking symmetries which were thought to be crucial
\citep[\eg][]{DrosselSchwabl:1992a} and later provided an ordering
principle
\citep{BihamMilshteinSolomon:1998,MilshteinBihamSolomon:1998,BihamMilshteinMalcai:2001,HughesPaczuski:2002,KarmakarMannaStella:2005}.
One reason why the subject of SOC remains contentious is the richness of
the results found. They often do \emph{not} fall clearly in one
universality class, in fact, even on the basis of extensive computer
simulations it is often not even possible to determine whether scaling
takes place at all, let alone the exponents
\citep[\eg][]{DornHughesChristensen:2001}.  The two computer models,
which so far display the clearest evidence for SOC, the Manna
\citep{Manna:1991a} and the Oslo Model \citep{ChristensenETAL:1996},
\emph{are} in the same universality class \citep{NakanishiSneppen:1997},
clearly not for trivial reasons. 

\begin{acknowledgement}
GP would like to thank the Kavli Institute for Theoretical Physics for
hospitality, and gratefully acknowledges the kind support by EPSRC Mathematics
Platform grant EP/I019111/1. NWW and SCC acknowledge grants from the Max Planck Society enabling Senior Visiting Scientist positions at MPIPKS in Dresden, and the warm and stimulating hospitality there of Holger Kantz and the members of his Time Series Analysis group. Over the years many people have freely shared their thoughts on SOC and other facets of complexity science with us and contributed greatly to our own view of the subject. We thank them all: in particular the late Per Bak and Leonardo Castillejo, and others including Gary Abel, Markus Aschwanden, John Beggs, Tom Chang, Kim Christensen, Giuseppe Consolini, Daniel Crow, Joern Davidsen, Richard Dendy, Deepak Dhar, Ronald Dickman, Mervyn Freeman, Martin Gerlach, Peter Grassberger, John Greenhough, Bogdan Hnat, Pierre Le Doussal, Sven Luebeck, Tony Lui, Sasha Milovanov, Nicholas Moloney, Maya Paczuski, Ole Peters, Mike Pinnock, Chris Rapley, Alan Rodger, George Rowlands, Kristoffer Rypdal, Martin Rypdal,  Cosma Shalizi, Surjal Sharma, Jouni Takalo, Vadim Uritsky, Alessandro Vespignani, Zoltan Voeroes, Clare Watt,  Dave Waxman and  Kay J{\"o}rg Wiese. \end{acknowledgement}

\newcommand{\bibconferencename}[1]{\emph{#1}}
\bibliography{ISSI_SOC_MathPhys}

\end{document}